\documentclass[]{aa}  

\DeclareUnicodeCharacter{FEFF}{}
\usepackage[]{natbib}
\usepackage[T1]{fontenc}
\usepackage{ae,aecompl}
\usepackage[dvipsnames]{xcolor}
\usepackage{psfrag,color}
\usepackage[]{inputenc}
\usepackage{graphicx}	
\usepackage{amsmath}	
\usepackage{amssymb}	
\usepackage{longtable}
\usepackage{subfigure}
\usepackage{adjustbox}
\usepackage{ulem}       
\usepackage{newtxmath} 
\usepackage{newtxtext}
\usepackage{pdflscape}
\usepackage{supertabular}
\usepackage{rotating}
\usepackage{footnote}
\usepackage{times}
\usepackage{chemmacros}
\usepackage{xspace}

\usepackage[colorlinks=true, urlcolor=blue, linkcolor=blue, citecolor=blue]{hyperref} 

\definecolor{lime}{HTML}{A6CE39}
\DeclareRobustCommand{\orcidicon}{
	\begin{tikzpicture}
	\draw[lime, fill=lime] (0,0) 
	circle [radius=0.16] 
	node[white] {{\fontfamily{qag}\selectfont \tiny ID}};
	\draw[white, fill=white] (-0.0625,0.095) 
	circle [radius=0.007];
	\end{tikzpicture}
	\hspace{-2mm}
}

\foreach \x in {A, ..., Z}{\expandafter\xdef\csname orcid\x\endcsname{\noexpand\href{https://orcid.org/\csname orcidauthor\x\endcsname}
			{\noexpand\orcidicon}}
}

\usepackage[]{ulem}  
\newcommand\redout{\bgroup\markoverwith
{\textcolor{red}{\rule[0.5ex]{2pt}{0.8pt}}}\ULon}
\defcitealias{2022Garcia-Rojas_mnras510}{GR22}
\defcitealias{richeretal22}{R22}

\newcommand{\ngc}{NGC~6153}
\newcommand{\Ne}{$n_{\rm e}$\xspace}
\newcommand{\Nev}[1]{$n_{\rm e}$ = #1$\,$cm$^{-3}$}
\newcommand{\Te}{$T_{\rm e}$\xspace}
\newcommand{\Tev}[1]{$T_{\rm e}$ = #1$\,$K}

\newcommand{\useaa}{} 
\newcommand{\ionic}[1]{$\,${\sc #1}}

\newcommand{\forb}[2]{[{#1}\ionic{#2}]}
\newcommand{\forbl}[3]{\forb{#1}{#2}$\,\lambda\,$#3\useaa}

\newcommand{\forbr}[4]{\forb{#1}{#2}$\,\lambda\,$#3/#4\useaa}

\newcommand{\perm}[2]{#1\ionic{#2}}
\newcommand{\perml}[3]{\perm{#1}{#2}$\,\lambda\,$#3\useaa}
\newcommand{\perms}[4]{\perm{#1}{#2}$\,\lambda\,$#3 + #4\useaa}

\newcommand{\hii}{\perm{H}{ii}}
\newcommand{\ha}{H$\alpha$\xspace} 
\newcommand{\hb}{H$\beta$\xspace}

\newcommand{\hi}{H\,{\sc i}}
\newcommand{\hei}{He\,{\sc i}\xspace}
\newcommand{\heii}{He\,{\sc ii}\xspace}
\newcommand{\nii}{N\,{\sc ii}}
\newcommand{\oii}{O\,{\sc ii}}

\def\vhel{\ifmmode{V_{{\rm HEL}}}\else{$V_{{\rm HEL}}$}\fi}
\def\vsys{\ifmmode{V_{\rm sys}}\else{$V_{\rm sys}$}\fi}
\def\kms{\ifmmode{~{\rm km\,s}^{-1}}\else{~km~s$^{-1}$}\fi}
\def\vlsr{\ifmmode{v_{\rm lsr}}\else{$v_{\rm lsr}$}\fi}

\newcommand{\pyneb}{\texttt{PyNeb}\xspace}

\newcommand{\forba}[3]{[\ion{#1}{#2}]~#3~\AA\xspace}
\newcommand{\alloa}[3]{\ion{#1}{#2}~#3~\AA\xspace}

\begin{document}



\title{MUSE spectroscopy of the high abundance discrepancy planetary nebula NGC\,6153 }

   \subtitle{}

    \author{V. G\'omez-Llanos\inst{1,2}{\orcidA{}},
    J. Garc\'{\i}a-Rojas\inst{1,2}{\orcidB{}},
    C. Morisset\inst{3,4}{\orcidC{}},
    H. Monteiro\inst{5}{\orcidD{}},
    D. Jones\inst{1,2,6}{\orcidE{}},
    R. Wesson\inst{7}{\orcidF{}},
    H.~M.~J. Boffin\inst{8}{\orcidG{}},
    R.~L~M. Corradi\inst{1,2,9}{\orcidH{}},
}
    \institute{Instituto de Astrof\'isica de Canarias, E-38205 La Laguna, Tenerife, Spain \email{vgomez.astro@gmail.com}
    \and
        Departamento de Astrof\'isica, Universidad de La Laguna, E-38206 La Laguna, Tenerife, Spain
    \and
        Instituto de Astronom\'ia (IA), Universidad Nacional Aut\'onoma de M\'exico, Apdo. postal 106, C.P. 22800 Ensenada, Baja California, M\'exico
    \and
        Instituto de Ciencias Físicas, Universidad Nacional Autónoma de México, Av. Universidad s/n, 62210 Cuernavaca, Mor., México 
    \and
        Instituto de F\'{\i}sica e Qu\'{\i}mica, Universidade Federal de Itajub\'a, Av. BPS 1303-Pinheirinho, 37500-903, Itajub\'a, Brazil 
    \and
        Nordic Optical Telescope, Rambla Jos\'e Ana Fern\'andez P\'erez 7, 38711, Bre\~na Baja, Spain 
    \and
        School of Physics and Astronomy, Cardiff University, Queen's Buildings, The Parade, Cardiff CF24 3AA, UK
    \and
        European Southern Observatory, Karl-Schwarzschild-Str. 2, 85738 Garching bei M\"unchen, Germany
    \and
        Gran Telescopio CANARIAS S.A., c/ Cuesta de San Jos\'e s/n, Bre\~na Baja, E-38712 Santa Cruz de Tenerife, Spain
        }

\authorrunning {G\'omez-Llanos et al.}
\titlerunning {MUSE spectroscopy of the PN NGC\,6153}
   \date{\today}

 
\abstract
{The abundance discrepancy problem in planetary nebulae (PNe) has long puzzled astronomers. NGC\,6153, with its high abundance discrepancy factor (ADF $\sim$10), provides a unique opportunity to study the chemical structure and ionisation processes within these objects.}   
{We aim to understand the chemical structure and ionisation processes in this high-ADF nebula by constructing detailed emission line maps and examining variations in electron temperature and density. This study also explores the discrepancies between ionic abundances derived from collisional and recombination lines, shedding light on the presence of multiple plasma components.}
{We used the MUSE spectrograph to acquire IFU data covering the wavelength range 4600$-$9300 \AA\ with a spatial sampling of 0.2 arcsec and spectral resolutions ranging from R = 1609 to R = 3506. We created emission line maps for 60 lines and two continuum regions. We developed a tailored methodology for the analysis of the data, including correction for recombination contributions to auroral lines and the contributions of different plasma phases. }
{Our analysis confirmed the presence of a low-temperature plasma component in NGC\,6153. We find that electron temperatures derived from recombination line and continuum diagnostics are significantly lower than those derived from collisionally excited line diagnostics. Ionic chemical abundance maps were constructed, considering the weight of the cold plasma phase in the \hi{} emission. Adopting this approach we found ionic abundances that could be up to 0.2 dex lower for those derived from CELs and up to 1.1 dex higher for those derived from RLs than in the case of a homogeneous \hi{} emission. The abundance contrast factor (ACF) between both plasma components was defined, with values, on average, 0.9 dex higher than the ADF. Different methods for calculating ionisation correction factors (ICFs), including state-of-the-art literature ICFs and machine learning techniques, yielded consistent results.}
{Our findings emphasise that accurate chemical abundance determinations in high-ADF PNe must account for multiple plasma phases. Future research should focus on expanding this methodology to a broader sample of PNe, with spectra deep enough to gather physical condition information of both plasma components, which will enhance our understanding of their chemical compositions and the underlying physical processes in these complex objects.}


   \maketitle
%



\section{Introduction \label{sec:intro}}

One of the unsolved questions in nebular astrophysics is whether the correct determination of the elemental abundances of elements heavier than helium in photoionised nebulae is provided by bright collisionally excited lines (CELs) or by faint recombination lines \citep[RLs, see e.g.][]{garciarojasesteban07}. In almost all objects studied so far, this abundance discrepancy is characterised for a given ion by the ratio of abundances obtained from RLs and CELs, which is called the abundance discrepancy factor (ADF). The ADF values are greater than 1, with a median value of about 2–3 for H II regions and the majority of planetary nebulae (PNe).\footnote{For an updated compilation of literature values of the ADF(O$^{2+}$), see \href{https://nebulousresearch.org/adfs/}{https://nebulousresearch.org/adfs/}}. 
Several scenarios have been proposed to explain this behaviour over the past decades. The most popular ones include the presence of temperature inhomogeneities in the gas \citep{peimbert67, torrespeimbertetal80}, chemical inhomogeneities in the gas \citep{torrespeimbertetal90, liuetal00, tsamispequignot05}, or deviations from a Maxwellian distribution of the energy of thermal electrons \citep[kappa distribution;][]{nichollsetal12}. Obtaining direct evidence to support any of these scenarios has proven to be a challenging task. However, observational evidence of the action of temperature fluctuations in photoionised nebulae has been found very recently, although it is restricted to {\hii} regions \citep{2023Mendez-Delgado_Natur618}.

Most of the scenarios proposed to explain the AD fail when the ADF is too high. This is the case with the increasing number of PNe exhibiting ADF values much higher than those observed in {\hii} regions, reaching values of 10 or more. However, several studies have proposed that the presence of at least two distinct plasma components in these PNe is the driver of the high ADFs found. Although this scenario was proposed more than thirty years ago to explain the general behaviour of AD in ionised nebulae \citep{torrespeimbertetal90}, it has only gained relevance in the last decade due to increasing observational evidence of the presence of two plasma components \citep[e.g.][]{corradietal15, garciarojasetal16, 2022Garcia-Rojas_mnras510, richeretal22}. In this scenario, one of the plasma components is very metal-rich (i.e. hydrogen-poor) and exhibits very different physical conditions compared to the main gas component that emits CELs. The origin of this metal-rich gas component is still a matter of debate, and several scenarios have been proposed, most of them linked to the fact that the majority of the high-ADF PNe appear to be associated with a close binary central star that has undergone a common envelope event \citep[see e.g.][]{corradietal15, jonesetal16, wessonetal18}.

In this context, in order to understand the physical processes that give rise to the phenomenon of high-ADF PNe, we need to refine the abundance determinations for each of the components. 
The advent of the Multi Unit Spectroscopic Explorer (MUSE) integral-field spectrograph  \citep{baconetal10} on the Very Large Telescope (VLT) has revolutionised the field of PN research, paving the way for high-spatial resolution studies that allow for detailed mapping of the physical and chemical properties on these objects. However, the first studies of the PNe NGC\,7009, NGC\,3132, and IC\,418 \citep{walshetal16, walshetal18, monrealiberowalsh20, monrealiberowalsh22} were done using the nominal mode, which covers the wavelength range 470$-$930\,nm that prevents the detection of the \oii\ RLs around $\sim$465\,nm and hence, not allowing the AD problem to be addressed. In order to circumvent this issue, \citet[][hereafter \citetalias{2022Garcia-Rojas_mnras510}]{2022Garcia-Rojas_mnras510} obtained deep observations of three high-ADF PNe using the MUSE extended mode, which covers the wavelength range $460-930$\,nm, and mapping, for the first time, the spatial distribution of the \oii\ RLs in this type of PN. \citetalias{2022Garcia-Rojas_mnras510} emphasised the importance of taking into account the contribution of recombination to the \forb{O}{ii} and \forb{N}{ii} auroral lines, which are crucial for a proper determination of the physical conditions of the warm gas, and the relative importance of hydrogen emission in both components, which is essential to properly compute the metal content of the cold gas component, as was pointed out in the photoionisation models presented by \citet{2020Gomez-Llanos_mnra497}. Regarding this last issue, \citetalias{2022Garcia-Rojas_mnras510} presented a methodology to compute the relative contribution of H to the H-poor gas component; however, given the lack of diagnostics to estimate the physical conditions of the H-poor gas in the three PNe studied, only rough estimates could be made in that study.  

\ngc\ is a bright, southern PN, with a relatively high ADF$\simeq$10 \citep{liuetal00}. The large amount of deep spectrophotometric data obtained for this object has made it the most studied PN in terms of the AD problem. Several observational techniques, from moderate-spectral-resolution long-slit spectroscopy scanning the entire nebular surface \citep{liuetal00} to high-spectral-resolution \'echelle spectroscopy \citep{mcnabbetal16,richeretal22}, have been used to analyse the physical conditions and chemical abundances in this object. 

\citet{tsamisetal08} attempted a 2D spectroscopic study in this PN using ARGUS/FLAMES at the VLT. The field-of-view was limited to a relatively small region of 11.5$\times$7.2 arcsec$^2$, covering from the central star to the south-eastern outskirts of the PN. These authors found that the ADF(O$^{2+}$) computed from the abundance ratio obtained from \perml{O}{ii}{4649} RL and \forbl{O}{iii}{4959} CEL reached values of $\sim$20 near the PN nucleus. 

Very recently, the most comprehensive and detailed spectroscopic study of NGC\,6153 has been developed by \citet[][hereafter \citetalias{richeretal22}]{richeretal22}. These authors took advantage of deep, high-spectral-resolution observations obtained with the UltraViolet Echelle Spectrograph (UVES) attached to UT2 of the VLT to simultaneously carry out studies of the chemical abundances and the kinematics of the gas. One of the most remarkable conclusions of the paper is the confirmation that emission from heavy-element optical RLs defines a different kinematic plasma component with a lower temperature and a higher density than the normal nebular plasma. Even more important is the fact that these authors managed to measure temperature-sensitive and density-sensitive {\oii} and {\nii} RLs, which allowed the computation of the physical conditions of the H-poor plasma component and, therefore, to break the degeneracies mentioned above.

In this paper we present deep MUSE spectrophotometry of the well-known PN NGC\,6153 where we follow a similar approach to that of \citetalias{2022Garcia-Rojas_mnras510} to study the spatial distribution of the physical conditions and ionic chemical abundances, taking advantage of the \citetalias{richeretal22} results to refine the contribution of the different plasma components to the emissivities of the lines emitted by the nebula.  

\section{Observations and data reduction}\label{sec:obs}

\ngc\ was observed with MUSE on the VLT, in seeing-limited mode, on the night of 6 to 7 July 2016. The log of the observations is provided in Table~\ref{Tab:Obs}. The sky conditions were reasonable with a seeing of around 1 arcsec and under thin clouds. 
The instrument was used in its Wide Field Mode with the natural seeing (WFM--NOAO) configuration. This provides a nearly contiguous 1\,arcmin$^2$ field of view with  0.2-arcsec spatial sampling. 
We used the extended mode of MUSE (WFM--NOAO-E), which covers the wavelength range $460-930$\,nm with an effective spectral resolution that increases from $R\sim1609$ at the bluest wavelengths to $R\sim3506$ at the reddest wavelengths.
The data are available at the ESO Science Archive under Prog.\ ID 097.D--0241(A) (PI: R.~L.~M. Corradi). 
Observations were obtained with dithering and a rotation of 90\degr{} between each exposure to remove artefacts during data processing. Given that the target is extended, we obtained two sky frames on each sequence (see Table~\ref{Tab:Obs}) to perform an adequate sky subtraction. A few short exposures were also taken to analyse the strongest emission lines.
The data reduction was done with {\tt esorex}, using the dedicated ESO pipeline \citep{weilbacheretal14, weilbacheretal20}. For each individual frame the pipeline performs bias subtraction, flat-fielding and slice-tracing, wavelength calibration, geometric corrections, illumination correction using twilight sky flats, sky subtraction (telluric absorption/emission lines and continuum) making use of the sky frames obtained between science exposures. In the final steps of the reduction process, differential atmospheric correction and flux calibration were performed.

\begin{table}
\caption{Log of the MUSE observations.}\label{Tab:Obs}
\begin{tabular}{ccccc}
\hline
UT Start & n & Exp &  Airm. & Seeing \\
& & (s) &  & ($"$) \\\hline
Target: \ngc & \multicolumn{4}{c}{Mode:  WFM--NOAO-E}\\ \hline
2016-07-07 01:57:13.657&1/9&10.0&1.04&0.7\\
 2016-07-07 01:59:22.152&2/9&60.0&1.04&0.68\\
 2016-07-07 02:02:21.913&3/9&450.0&1.039&0.81\\
 2016-07-07 02:11:16.062&4/9$^{\rm a}$&180.0&1.039&0.96\\
 2016-07-07 02:16:23.337&5/9&450.0&1.039&0.86\\
 2016-07-07 02:25:52.667&6/9&450.0&1.04&0.83\\
 2016-07-07 02:34:47.108&7/9$^{\rm a}$&180.0&1.042&1.07\\
 2016-07-07 02:39:55.206&8/9&450.0&1.044&0.91\\
 2016-07-07 02:49:25.123&9/9&450.0&1.048&0.90\\ \hline
\hline
\end{tabular}
\begin{description}
\item $^{\rm a}$ {Sky frames were taken 5 arcmin away from the object to ensure that there was no nebular contamination.}
\end{description}
\end{table}

\begin{figure}
	\centering
	
	{\includegraphics[width=9.5cm]{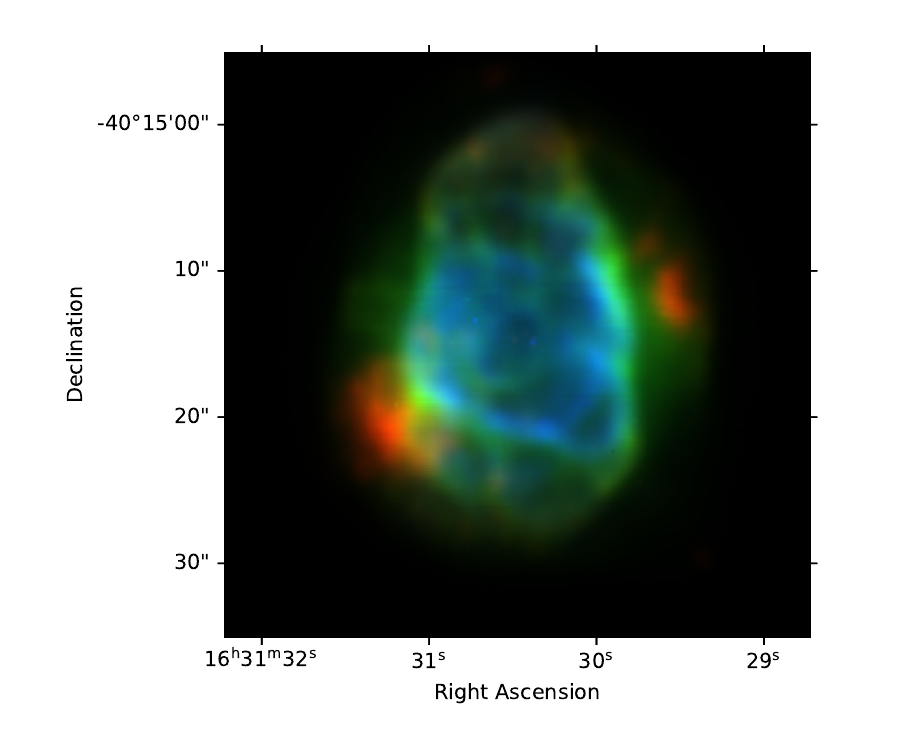}}
	\caption{Composite RGB image of the MUSE field of view for \ngc. {\perml{He}{ii}{4686}} is shown in blue, {\hb} in green and an average of {\forbl{N}{ii}{6548}} and {\forbl{S}{ii}{6716}} in red. Intensity scale is linear. North is up, and east to the left. \label{fig:rgb}}
\end{figure}

To speed-up computations, we trimmed the original data cubes to the central 40 arcsec (200 $\times$ 200 spaxels) which is enough to cover the main nebular emission shells. 
In Fig.~\ref{fig:rgb}, we present a RGB composite image of \ngc\ observed with MUSE, the ionisation stratification can be observed with the different colours: red for an average of {\forbl{N}{ii}{6548}} and {\forbl{S}{ii}{6716}}, green for \hb, and blue for {\perml{He}{ii}{4686}}. 

As in \citetalias{2022Garcia-Rojas_mnras510}, we checked for possible saturation of the brightest lines with a preliminary inspection of the data cubes with {\sc qfitsview} \citep{ott12}, finding that \forbl{O}{iii}{5007} and \ha lines were slightly saturated in the brightest zones. We therefore avoid using \forbl{O}{iii}{5007} measurements and to minimise the effects of saturation in \ha, we followed the same procedure as in \citetalias{2022Garcia-Rojas_mnras510} to compute the extinction coefficient. 

\section{Emission line measurements}\label{sec:lines}

We constructed flux maps for 60 emission lines (30 RLs and 30 CELs). The complete list of emission lines is shown in Table \ref{tab:int_fluxes}. All the fluxes and associated errors were measured as described in Sect. 3 of \citetalias{2022Garcia-Rojas_mnras510}. In addition, we constructed continuum maps at 8100 \AA\ and 8400 \AA, using the automated line-fitting algorithm {\sc ALFA} \citep{wesson16}. To only consider high signal-to-noise ratio observations, we define the following mask: F(H$\beta$)$>$0.005 $\times$ F(H$\beta$)$_{max}$, that we apply to all emission line maps and their subsequent analysis. 

In Fig.~\ref{fig:line_fluxes} we present the unreddened fluxes of the most relevant emission lines, sorted by atomic mass and ionisation potential (IP) of the ion. The RLs of metals present a similar spatial distribution regardless of the element and IP of the emitting ion (see \perml{C}{ii}{6462}, \perml{N}{ii}{5679}, \perml{N}{iii}{4641}\footnote{Line \perml{N}{iii}{4640.64} is blended with \perml{O}{ii}{4641.81}, and possibly with \perml{O}{ii}{4638.86} and \perml{N}{iii}{4641.85}.}, \perml{O}{i}{7773+}, and \perml{O}{ii}{4649}\footnote{Corresponds to \perms{O}{ii}{4649.13}{4650.85}, and possibly affected by a blend with \perml{C}{iii}{4650.25}. We will check the effect of this blend in chemical abundance maps in Sect.~\ref{sec:ionic_ab_maps}.}), unlike the emission maps of CELs where there is a clear ionisation structure when comparing the emission of ions with different IP of the same or different element (e.g. \forbl{N}{i}{5198}, \forbl{N}{ii}{6548}, \forbl{Ar}{iii}{7136}, \forbl{Ar}{iv}{4740}, \forbl{Ar}{v}{7005}). The emission line \forbl{O}{ii}{7330+}, which shows a morphology similar to the RLs of metals in the inner parts of the nebula, has a non-negligible recombination contribution that is estimated and subtracted from the total emission in Sec.\ref{sec:corr_rec}.

We present a spatially resolved map of a neutron-capture emission line (\forb{Kr}{iv} at 5867.74 \AA) (see Fig.~\ref{fig:line_fluxes}). To our knowledge, this emission has been previously spatially mapped only in the PN IC\,2165 \citet{otsuka22}. The \forbl{Kr}{iv}{5867.74} line emission map clearly resembles the spatial distribution of emission lines with a similar ionisation potential range, such as \forbl{O}{iii}{4959}. 

\section{Mapping the physical and chemical properties of \ngc}\label{sec:phys_cond_ionic}

We use the v1.1.18 of \pyneb \citep{luridianaetal15} for the emission line analysis, including 150 Monte Carlo simulations for the uncertainties estimations, following the protocol described in Sec.~3.2 of \citetalias{2022Garcia-Rojas_mnras510}. These Monte Carlo simulations allow us to generate uncertainty maps for each quantity computed in this section. In the following section, we only briefly describe the pipeline used to obtain the electron temperature and density, and the ionic abundances, and refer to \citetalias{2022Garcia-Rojas_mnras510} for details.

\subsection{Presence of a metal-rich, cold plasma phase}
\label{sec:coldregion}

As already mentioned in Sect.~\ref{sec:intro}, in PNe demonstrating high ADFs (greater than 10), the presence of a high-metallicity central cold phase of gas is now well established, particularly owing to the efforts made by the community in obtaining spatially resolved observations of these types of objects \citep{wessonetal03, corradietal15, garciarojasetal16, 2022Garcia-Rojas_mnras510, richeretal22}. In the subsequent sections of this paper, we explain how we computed the ionic and total abundances accounting for the existence of two phases of gas: one being the ``classical'' warm shell emitting the CELs, and the other, more centrally emitted, metal-rich (or H-poor, with the difference lying in the He abundance; see below) and cold, as metal IR lines efficiently cool the gas. This second component serves as the primary source of the metal RLs.

Theoretical studies concerning the impact of this cold component have been published, for instance, by \citet{2002Pequignot_12, 2011Yuan_mnra411, 2020Gomez-Llanos_mnra497}, and \citet{2023Morisset_arXiv}. In the latter paper, a theoretical framework is outlined to consider the emission emanating from each region when determining the metal abundances.

In the following sections, we will operate under the assumption that the emission lines resulting from collisional excitation originate exclusively from the warm region, while those resulting from electron recombination to a metal ion originate exclusively from the cold region. Some lines may arise from both processes (e.g. auroral lines); this will be discussed in Section~\ref{sec:corr_rec}.

Following the convention of \citet{2020Gomez-Llanos_mnra497}, we will adopt the acronym ``ACF'' for the abundance contrast factor. This is because, under the assumption of the presence of the cold region, there is no longer any abundance discrepancy, but rather two distinct components with different abundances.

\subsection{Extinction correction}\label{sec:extict}

\begin{figure}
	\centering
	\includegraphics[width=8.5cm]{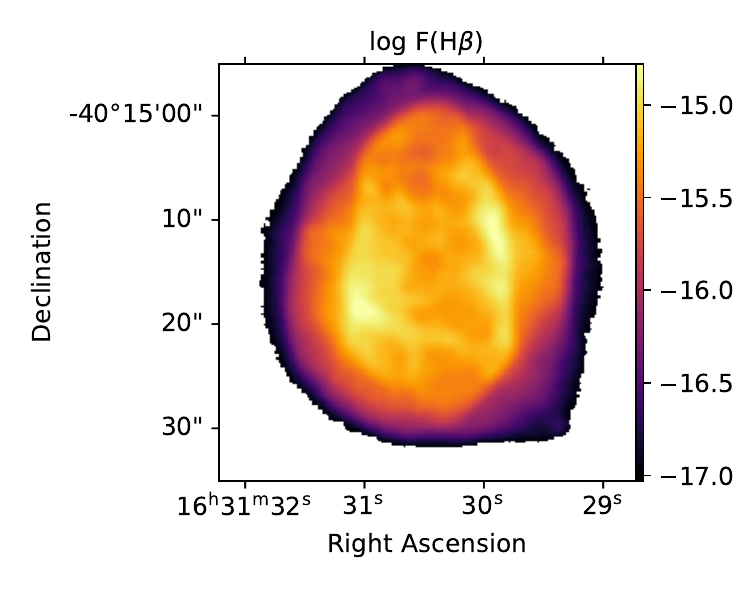}
	\caption{Spatially resolved map of the \hb\ measured flux (in units of erg cm$^{-2}$s$^{-1}$\AA$^{-1}$) in logarithmic scale.\label{Fig:fHb}}
\end{figure}

The observed flux map of {\hb} is presented in Fig.~\ref{Fig:fHb}, in logarithmic scale. From this bright emission line, we can distinguish the main nebular shell with two bright regions located north-west and south-east of the central star and a diffuse envelope.

\begin{figure}
    \centering
    \includegraphics[width=8.5cm]{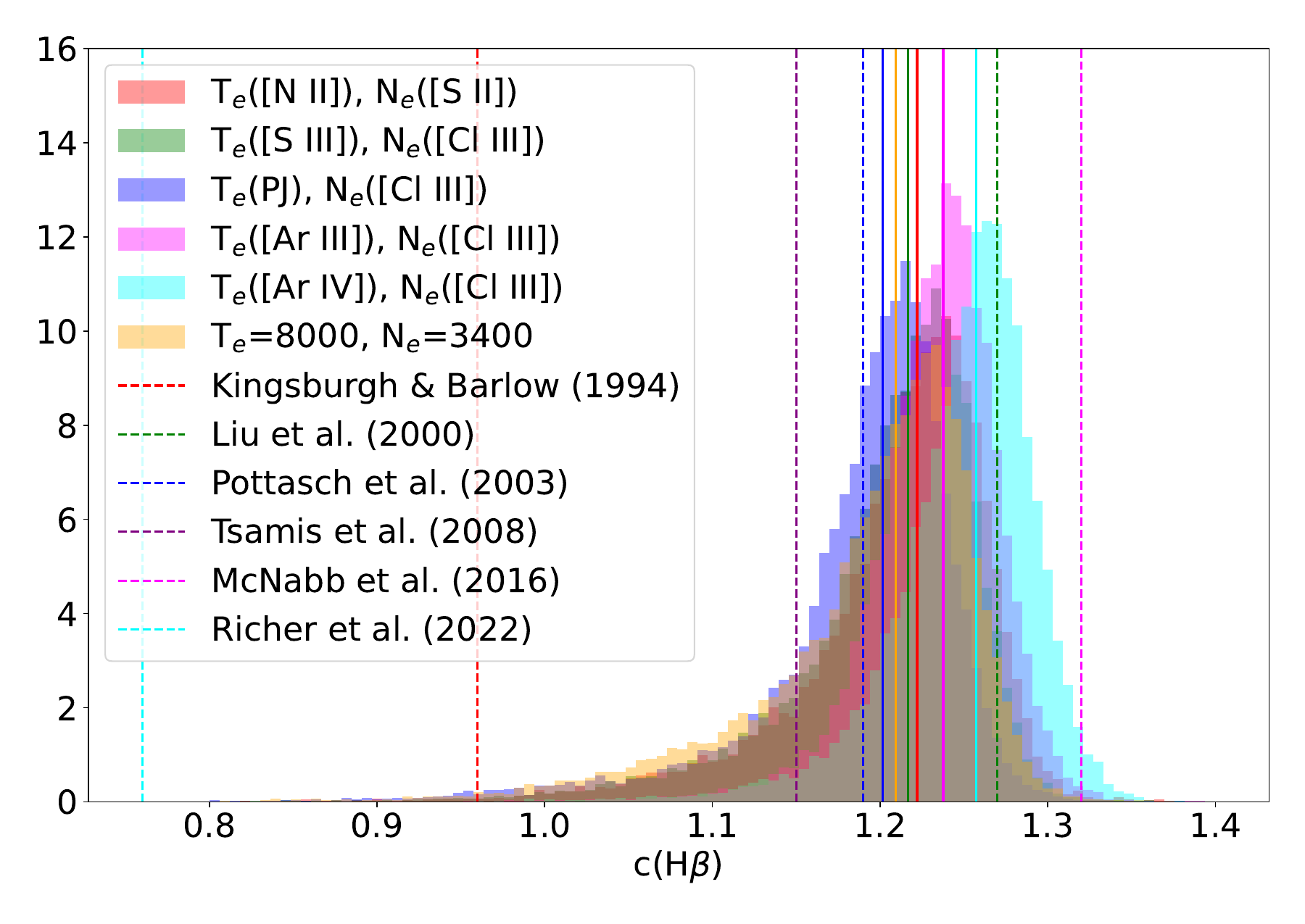}
    \caption{c(\hb) distributions obtained using different combinations of spatially resolved values of \Ne and \Te as well as based on uniform \Nev{3,400} and \Tev{8,000}. Vertical lines correspond to several estimates from the literature (see text).}
    \label{fig:cHb_dist}
\end{figure}
\begin{figure}
	\centering
	\includegraphics[width=9.0cm]{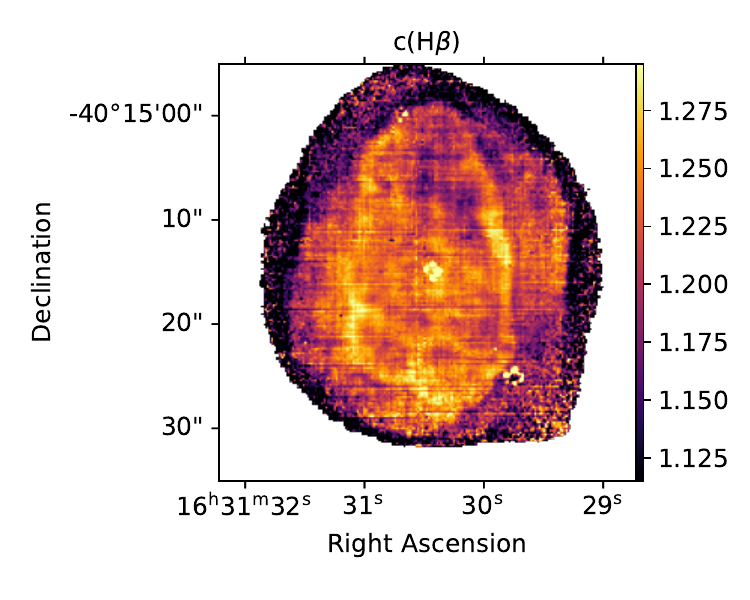}	\caption{Extinction map in logarithmic scale: c(\hb) is obtained from the median of the extinction maps based on four Paschen lines (P9 to P12) and H$\alpha$, normalised to H$\beta$.\label{Fig:cHb}}
\end{figure} 

The extinction correction c(\hb) is determined with \pyneb, using the pixel by pixel median of the values obtained from \ha/\hb, P9/\hb, P10/\hb, P11/\hb, and P12/\hb, by comparing observed values with theoretical values from \citep{storeyhummer95}, and using the extinction law defined by \cite{fitzpatrick99} with R$_V$=3.1. The theoretical line ratios have been obtained using \Nev{3,400} and \Tev{8,000}. These values correspond to what is found later in the process from the {\forbr{Cl}{iii}{5518}{5538}} and {\forbr{S}{iii}{6312}{9069}} line ratios, respectively (see Sect.~\ref{sec:phys_cond}).

\cite{uetamasaaki21} state that for spatially resolved observations, the spatial variations in c(\hb) cannot be recovered using a uniform (\Ne, \Te). An iteration is performed between the spaxel-by-spaxel values of \Ne and \Te determined later in the process, and c(\hb), for different diagnostic line ratios.
We calculate the distribution of c(\hb) using five different combinations of spatially resolved \Ne and \Te. In Fig.~\ref{fig:cHb_dist} we present these five c(\hb) distributions as well as the c(\hb) based on uniform \Nev{3,400} and \Tev{8,000}, and several estimates from the literature\footnote{\citet{kingsburghbarlow94}, \citet{liuetal00}, \citet{2003Pottasch_aap409}, \citet{tsamisetal03}, \citet{mcnabbetal16}, and \citetalias{richeretal22}}. The c(\hb) distributions are very consistent with each other and with the values from the literature, except for \citet{kingsburghbarlow94} and \citetalias{richeretal22}, which show much lower extinctions, especially \citetalias{richeretal22}, whose c(\hb) is completely outside the distributions computed here. 
Given the similarities of the results presented in Fig.~\ref{fig:cHb_dist} and to avoid adding noise to all emission maps, we choose to use c(\hb) obtained from uniform values for \Te and \Ne instead of spatially resolved maps.

The resulting map of c(\hb) is shown in Fig.~\ref{Fig:cHb}. We notice spatial variations in the extinction map of the order of approximately $\pm$6 per cent around a median value of 1.2, with a pattern similar to the spatial distribution of F(\hb) (see Fig.~\ref{Fig:fHb}). This indicates that part of the extinction is due to dust located inside the nebula. A similar behaviour is found for NGC~7009 \citep{walshetal16}, where c(\hb) roughly follows F(\hb). These authors compute the extinction map based on MUSE observations of the same Balmer and Paschen emission line ratios, correcting \ha and \hb from the \perm{He}{ii} Pickering lines and following an iterative process to recalculate the theoretical line ratios based on the latter estimations of \Te(\forb{S}{iii}) and \Ne(\forb{Cl}{iii}). The variations in c(\hb) found for NGC~7009 are much higher (more than 30 per cent) than what we find here for \ngc. 

The presence of the cold component described in Sec.~\ref{sec:coldregion} can affect the determination of the extinction correction because it affects the theoretical value to which the observed value is compared to obtain the correction. Anticipating the results presented in the following sections, we take both components into account using Eq.~6 from  in \citet{2023Morisset_arXiv} and compute the theoretical value for \ha/\hb as being between 2.90 and 3.03, the minimum being the case of a single region at \Nev{3,400} and \Tev{8,000}, and the maximum corresponding to spaxels where the contribution of the cold region is maximum. These high values in the theoretical \ha/\hb would reduce the determination of c(\hb) from, for example, 1.2 to 1.14 at maximum. This would, for example, reduce the \forb{S}{iii} (\forb{N}{ii}) electron temperatures from 8,300~K to 8,247~K (8,278~K respectively). Therefore, to avoid adding noise to the data, we do not apply any correction for the presence of the cold region.

\subsection{Correction for recombination contribution}
\label{sec:corr_rec}

In the case of the auroral lines {\forbl{N}{ii}{5755}} and {\forbl{O}{ii}{7320+30}}, it is well known that they are also produced by recombination, an emission process favoured at low temperatures. In the bulk of {\hii} regions and PNe, this contribution is relatively small, and the correction is much smaller than the typical uncertainties. However, this is not the case for high-ADF PNe, where  extremely strong recombination emission of both {\nii} and {\oii} lines from an hypothetical cold emission plasma can significantly contribute to the observed flux of these lines \citep[see][]{gomezllanosetal20}. Following the procedure described in Sec.~5.2 of \citetalias{2022Garcia-Rojas_mnras510}, we correct the observed emission of {\forbl{N}{ii}{5755}} and {\forbl{O}{ii}{7320+30}} from the contribution due to recombination using {\perml{N}{ii}{5679}} and {\perml{O}{ii}{4649+50}} respectively by applying the following relations:

\begin{equation}
    I(7320+30)_{\rm corr} = I(7320+30) - \frac{j_{7320+30}(T_{\rm e}, n_{\rm e})}{j_{4649+50}(T_{\rm e}, n_{\rm e})}\times I(4649+50)
\label{eq:oii_rec}
\end{equation}

\begin{equation}
    I(5755)_{\rm corr} = I(5755) - \frac{j_{5755}(T_{\rm e}, n_{\rm e})}{j_{5679}(T_{\rm e}, n_{\rm e})}\times I(5679)~,
\label{eq:nii_rec}
\end{equation}

\noindent
where $j_{\nu}({T_e}, {N_e})$ are the recombination emissivities for each line. 

\begin{figure}
	\centering
	\includegraphics[width=8.cm]{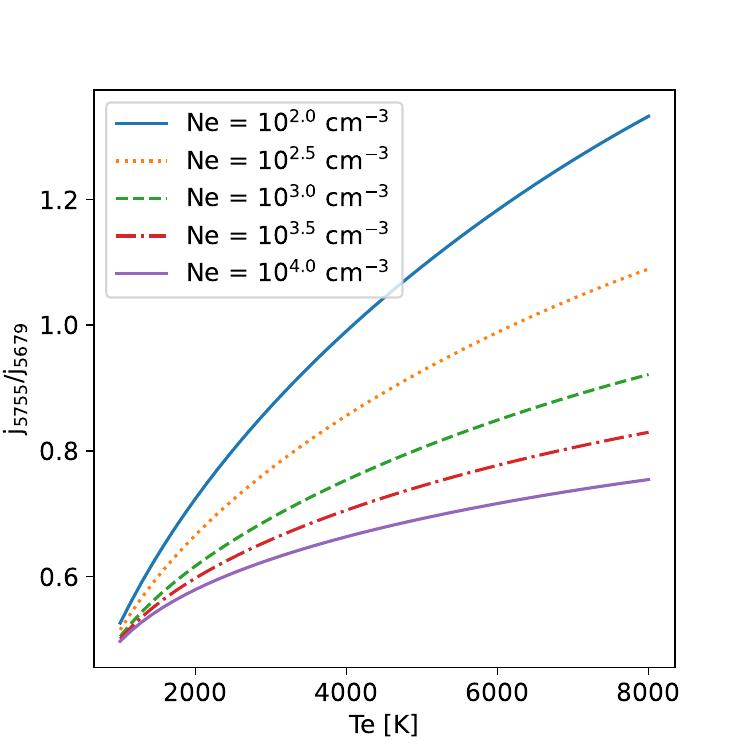}
	\caption{Relationship between the emissivity due to the recombination of \forba{N}{ii}{5755} and the emissivity of \alloa{N}{ii}{5679}. The ratios obtained for $n_{\rm e}$ lower than 100 cm$^{-3}$ are indiscernible.\label{Fig:niirec}}
\end{figure}

The relation between $j_{5755}$(\Te, \Ne)$ / j_{5679}$(\Te, \Ne) and \Te, for different values of \Ne is shown in Fig.~\ref{Fig:niirec}.  The emissivities have been obtained using \pyneb and the atomic data from \citet{pequignotetal91} and from \citet{fangetal11}  for $j_{5755}$ and $j_{5679}$, respectively. 

\begin{figure*}
    \centering
    \includegraphics[scale= 0.75]{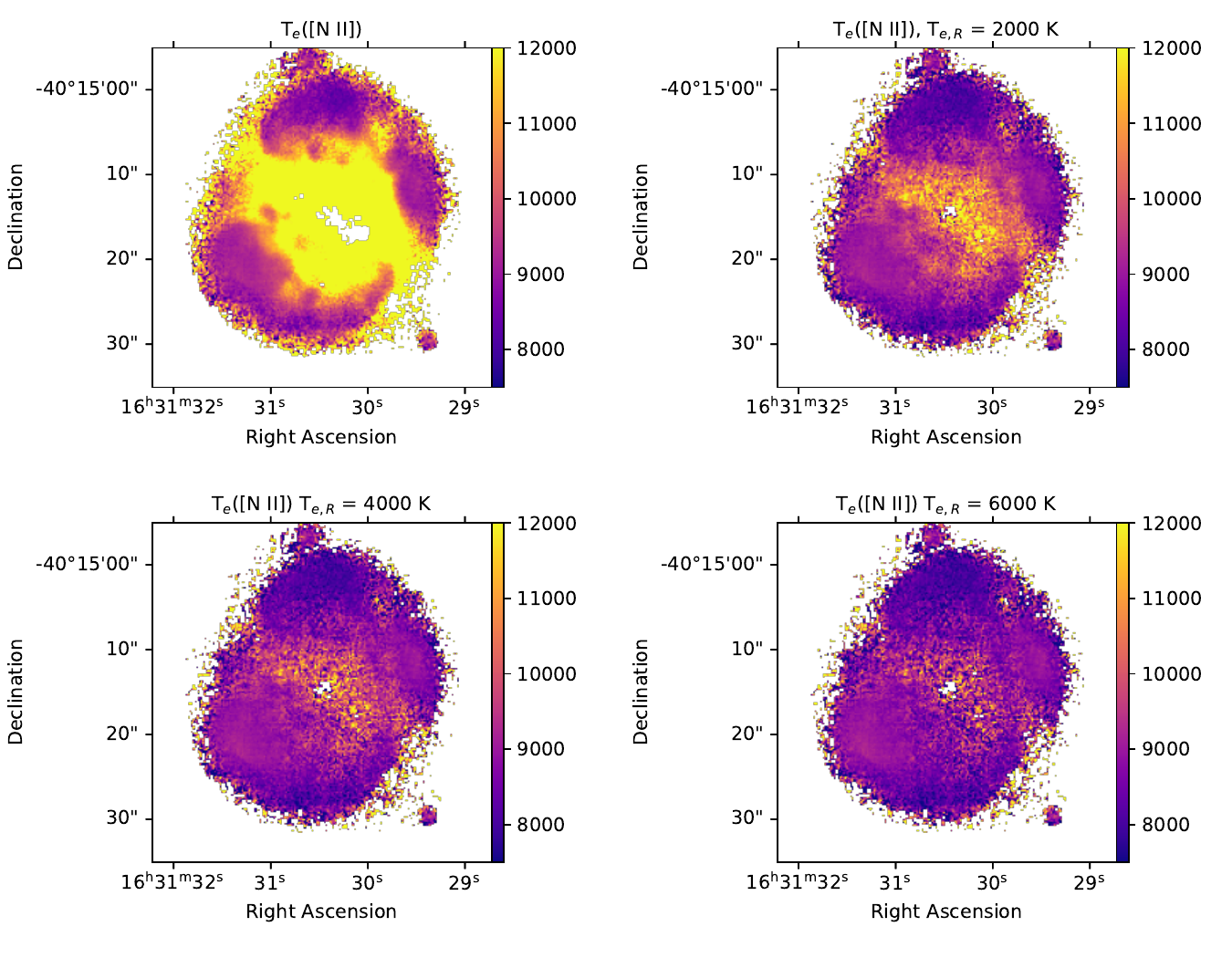}
    \caption{Map of the \forb{N}{ii} 5755/6584 electron temperature. The upper-left panel shows the result obtained without correcting from the recombination. The following panels are showing the results obtained when the \forbl{N}{ii}{5755} line is corrected for recombination using Eq.~\ref{eq:nii_rec} and different values of the electron temperature to compute the recombination line emissivities. The value of this temperature is indicated in each panel.}
    \label{fig:nii_te_corr_rec}
\end{figure*}

\begin{figure*}
    \centering
    \includegraphics[scale = 0.7]{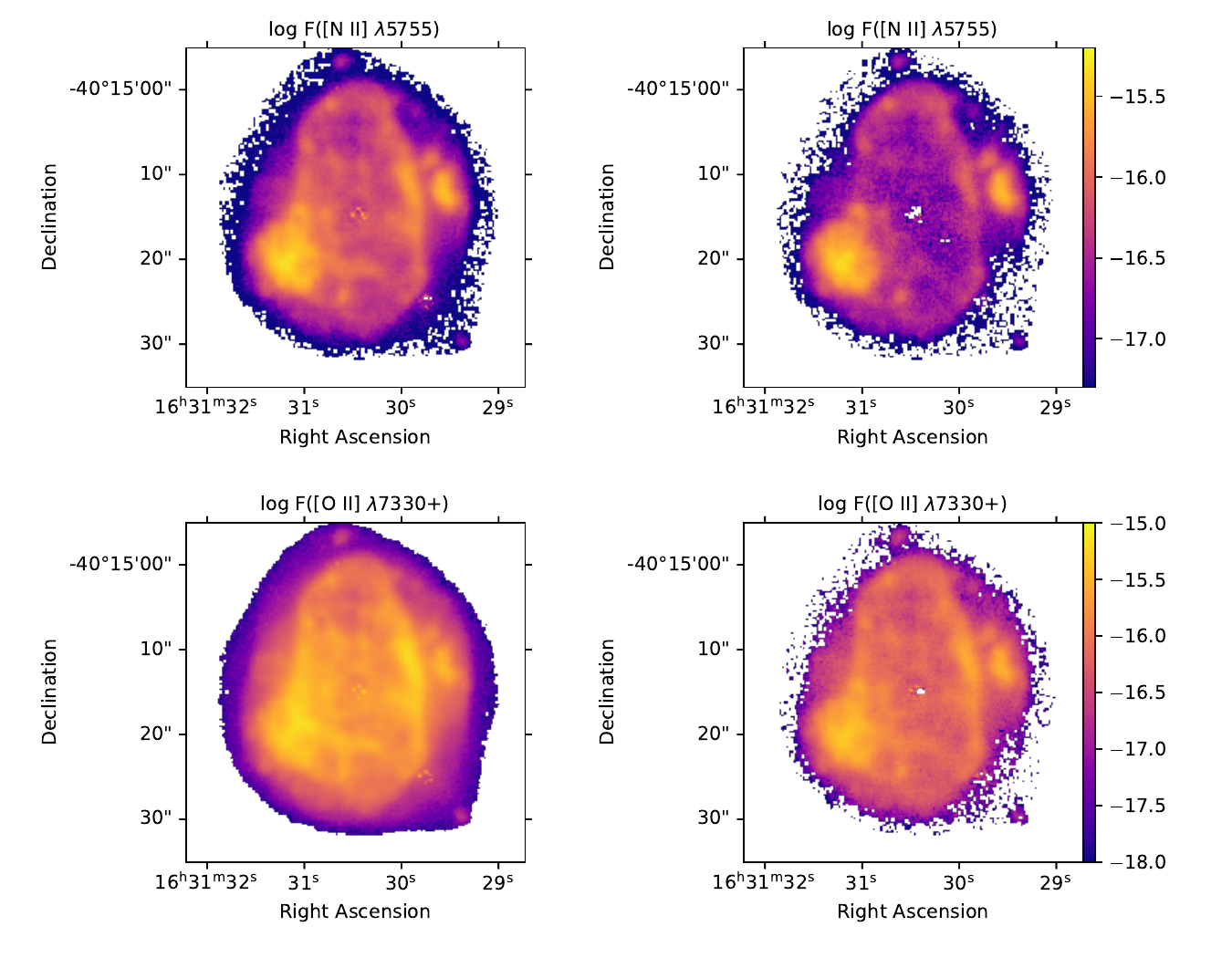}
    \caption{Emission maps of the two auroral lines \forbl{N}{ii}{5755} (upper panels) and \forbl{O}{ii}{7330+} (lower panels), without and with correction from recombination (left and right panels, respectively). The correction is obtained using Eqs.~\ref{eq:nii_rec} and \ref{eq:oii_rec}, adopting \Tev{6,000} and \Nev{10,000}. The flux is expressed in units of erg cm$^{-2}$s$^{-1}$\AA$^{-1}$ and in logarithmic scale.}
    \label{fig:nii_oii_corr_rec}
\end{figure*}

We explore the effect of correcting for the contribution of recombination on the determination of the temperature from the {\forbr{N}{ii}{5755}{6548}} line ratio (Eq.~\ref{eq:nii_rec}), using different values for the electron temperature of the region emitting the RLs: \Tev{2,000, 4,000 and 6,000}, and \Nev{10,000}. The resulting maps, as well as the result obtained without any correction, are shown in Fig.~\ref{fig:nii_te_corr_rec}. The map obtained with \Tev{6,000} to calculate the recombination correction is the most uniform: we will then adopt \Tev{6,000} for the correction. As explained in detail in \citetalias{2022Garcia-Rojas_mnras510}, the exact value of \Te is not strictly related to the temperature of the region that emits the lines, but should rather be considered as a tuning parameter. The fact that it is not the same as the value of 2,000~K obtained by \citetalias{richeretal22} from \perm{O}{ii}\ lines may point to the uncertainties in the line emissivities in Eq.~\ref{eq:nii_rec} and shown in Fig.~\ref{Fig:niirec}, especially given that the ratio of emissivities is obtained by combining two different sources of atomic data. Some fluorescence mechanism may also compromise these emissivities (See Sec.~3.3 of \citetalias{richeretal22}). 
We show in Fig.~\ref{fig:nii_oii_corr_rec} the effect of the correction on the auroral line emissions. Both maps on the left are not corrected, while the two maps on the right are corrected using \Tev{6,000}. The central part of the nebula is the most affected by the correction (this is where the recombination contamination dominates).

Our results differ significantly from what was obtained by \citet{liuetal00} and \cite{mcnabbetal16}. \citet{liuetal00} estimated varying recombination contributions to \forbl{N}{ii}{5755} between 7 and 64 per cent, depending on which N$^{2+}$/H$^+$ ratios were assumed, those given by far-infrared \forb{N}{iii} 57-$\mu$m line or by optical \perm{N}{ii} RLs, respectively. These corrections yield \Te(\forb{N}{ii}) between 9910\,K and 7110\,K, which are either too high or too low when compared with the value obtained from our integrated spectra following the methodology described above (see Sect.~\ref{sec:int_phys_cond}). On the other hand, \citet{mcnabbetal16} estimated a very small recombination contribution to \forbl{N}{ii}{5755} of only $\sim$1 per cent, leading to a \Te(\forb{N}{ii}) which is $\sim$1800\,K higher than our estimate from the integrated spectrum. From the integrated spectrum of NGC\,6153, we have computed a 40 per cent recombination contribution to \forbl{N}{ii}{5755} (see Sect.~\ref{sec:integrated}). 

Regarding the recombination contribution to the {\forbl{O}{ii}{7320+30}} lines, \citet{liuetal00} estimated that all the measured flux from these lines was due to recombination in NGC\,6153. Our analysis points to  different conclusions, as the flux maps of {\forbl{O}{ii}{7330+}} and \perml{O}{ii}{4649+} lines shown in Fig.~\ref{fig:line_fluxes} clearly reveal that the emission of both lines is not co-spatial. Following the described methodology, we estimated a recombination contribution of 51 per cent to the intensity of the {\forbl{O}{ii}{7320+30}} CELs in the integrated spectrum of NGC\,6153, which seems much more reasonable given the observed emission line maps. 

\begin{figure*}
    \sidecaption
    \includegraphics[width=12cm]{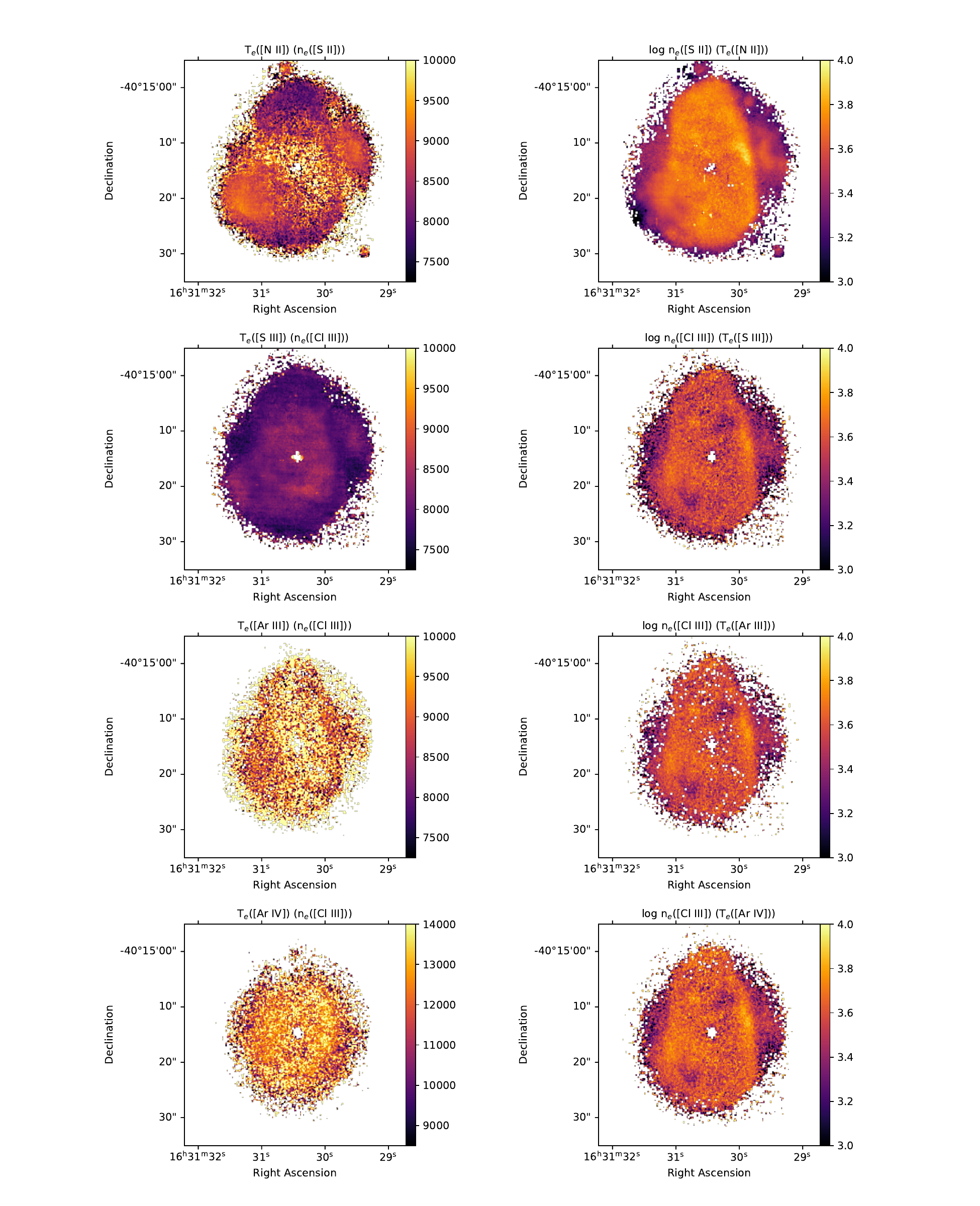}
    \caption{Physical condition maps obtained using different pairs of \Te-\Ne diagnostics. The \Te(\forb{N}{ii}) map shown in the first row has been corrected for recombination contribution.  
    \label{fig:te_ne}}
\end{figure*}

\begin{figure*}
    \centering
    \includegraphics[scale = 0.47]{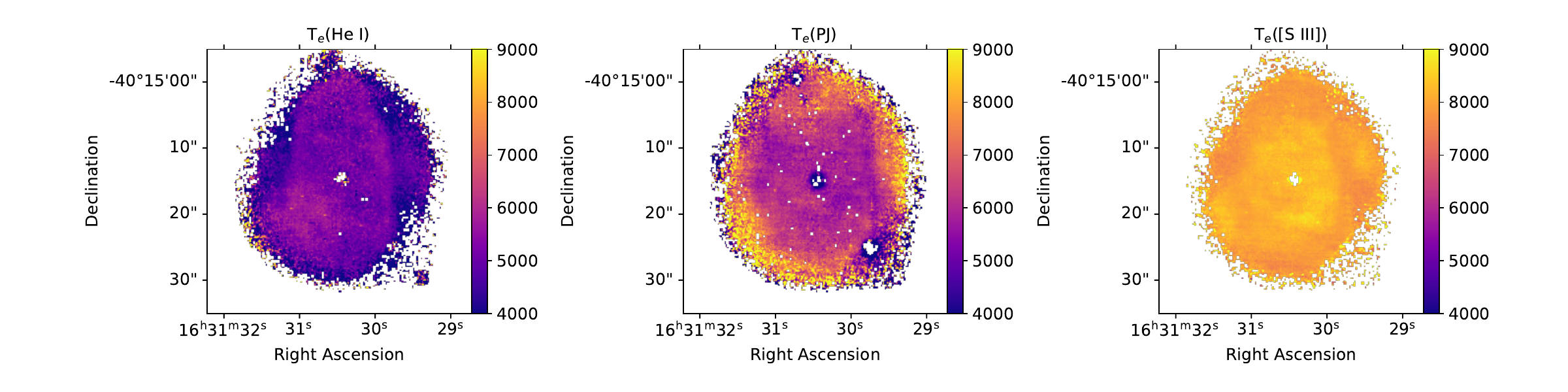}
    \caption{Temperature maps obtained from \hei lines (left panel), from the ratio of the Paschen jump to {\hi} P9 (central panel), and from the \forb{S}{iii} $\lambda$6312/$\lambda$9069 CEL ratio (right panel). The \Te scale is the same in all panels to better appreciate the differences.}
    \label{fig:Te_HI_PJ_S3}
\end{figure*}

\subsection{Physical conditions}
\label{sec:phys_cond}

\begin{table}
\caption{Atomic data sets used for the CELs and ORLs. \label{tab:atomic_data}}
\resizebox{\columnwidth}{!}{%
\begin{tabular}{lcc}
\hline
\multicolumn{3}{c}{CELs} \\
Ion & Transition Probabilities & Collision Strengths \\
\hline
O$^{0}$   &  \citet{wieseetal96} & \citet{bhatiaetal95}\\
O$^{+}$   &  \citet{wieseetal96} & \citet{kisieliusetal09}\\
O$^{2+}$  &  \citet{frosefischertachiev04} & \citet{storeysochi14}\\
          &  \citet{storeyzeippen00} & \\
N$^{+}$   &  \citet{frosefischertachiev04} & \citet{tayal11}\\
S$^{+}$   &  \citet{rynkunetal19} & \citet{tayal10}\\
S$^{2+}$  &  \citet{froesefischeretal06} & \citet{tayalgupta99}\\
Cl$^{2+}$ &  \citet{rynkunetal19} & \citet{butlerzeippen89}\\
Cl$^{3+}$ &  \citet{mendozazeippen82a} & \citet{butlerzeippen89}\\
 &  \citet{kaufmansugar86}  & \\
Ar$^{2+}$ &   \citet{munozburgosetal09}  & \citet{munozburgosetal09}\\
Ar$^{3+}$ &   \citet{rynkunetal19}  & \citet{ramsbottombell97}\\
Ar$^{4+}$ &   \citet{kaufmansugar86}  & \citet{galavisetal95}\\
 &  \citet{mendozazeippen82a}  & \\
\hline
\multicolumn{3}{c}{ORLs} \\
Ion & \multicolumn{2}{c}{Effective Recombination Coefficients}
 \\
\hline
H$^{+}$   & \multicolumn{2}{c}{\citet{storeyhummer95}} \\
He$^{+}$   & \multicolumn{2}{c}{\citet{porteretal12,porteretal13}} \\
He$^{2+}$   & \multicolumn{2}{c}{\citet{storeyhummer95}} \\
O$^{+}$   & \multicolumn{2}{c}{\citet{pequignotetal91}} \\
O$^{2+}$   & \multicolumn{2}{c}{\citet{storeyetal17}}\\
C$^{2+}$   & \multicolumn{2}{c}{\citet{daveyetal00}} \\
N$^{2+}$   & \multicolumn{2}{c}{\citet{fangetal11, fangetal13}}\\
\hline
\end{tabular}
}
\end{table}

The electron temperature and density, \Te and \Ne, are obtained by looking for the crossing point in 2D diagnostic diagrams combining, for example, the {\forbr{N}{ii}{5755}{6548}} and the {\forbr{S}{ii}{6716}{6731}} line ratios. This is carried out with the \pyneb library and is described in more detail in \citetalias{2022Garcia-Rojas_mnras510}, including the description of the artificial neural network used to accelerate the \texttt{Diagnostic.getCrossTemDen} method. The atomic data used are presented in Table~\ref{tab:atomic_data}, and pre-defined in \pyneb in the 'PYNEB\_23\_01' dictionary.

The diagnostics used to derive \Te and \Ne are listed in Table~\ref{tab:te_ne_diags}, grouped by the ionisation of the emitting region. In the case of the very high ionisation zone, the  \forbr{Ar}{iv}{4740}{4711} density diagnostic is not used due to the very noisy map of \forbl{Ar}{iv}{4711}, and \forbr{Cl}{iii}{5518}{5538} is preferred. In Fig.~\ref{fig:te_ne} we show the maps corresponding to \Te and \Ne obtained from each of these diagnostic pairs. The \Te(\forb{N}{ii}) map has been corrected from recombination contribution as explained in Sect.~\ref{sec:corr_rec}.

We also obtain \Te from \hei RLs and Paschen jump (PJ), following the methodology described in \citetalias{2022Garcia-Rojas_mnras510}. The corresponding maps are shown in Fig.~\ref{fig:Te_HI_PJ_S3}, as well as \Te(\forb{S}{iii}) for comparison. It is important to emphasize that the \Te{} maps obtained from both {\hei} lines and PJ are heavily affected by the presence of the low temperature component (described in Sec.~\ref{sec:coldregion}) with significant emission in these lines. 

\begin{table}
    \caption{Diagnostic line ratios used to compute the electron temperature and density simultaneously.}
    \centering
    \begin{tabular}{ccc}
    \hline
    Ionisation zone & Temperature diagnostic & Density diagnostic\\
    \hline
     Low  & \forbr{N}{ii}{5755}{6548}  & \forbr{S}{ii}{6716}{6731} \\
     High 
          & \forbr{S}{iii}{6312}{9069} & \forbr{Cl}{iii}{5518}{5538} \\
          & \forbr{Ar}{iii}{5192}{7136} & \forbr{Cl}{iii}{5518}{5538} \\
Very High 
          & \forbr{Ar}{iv}{4740}{7170} & \forbr{Cl}{iii}{5518}{5538} \\
    \hline
    \end{tabular}
    
    \label{tab:te_ne_diags}
\end{table}

\subsection{Ionic chemical abundance maps}
\label{sec:ionic_ab_maps}

Ionic chemical abundance maps are constructed only for the emission lines with the highest signal-to-noise ratio for each ion detected in the NGC\,6153 spectrum.
To determine the ionic abundances from these emission lines, we have to define the \Te and \Ne values to be used in computing the line emissivities and the \hb\ emissivity. The case of He$^+$/H$^+$ and He$^{2+}$/H$^+$ is relatively complicated, as there is no single plasma component emitting helium, but rather two different components: warm and cold. We will discuss them separately in Sect.~\ref{sec:ionic_ab_maps_he}. For the RLs of heavy elements, we use the \Te and \Ne values of 2,000~K and 10,000~cm$^{-3}$ obtained by \citetalias{richeretal22}.
For the CELs, we use the \Te$-$\Ne values derived from the {\forbr{N}{ii}{5755}{6548}$-$\forbr{S}{ii}{6716}{6731}} for IP<17~eV, and {\forbr{S}{iii}{6312}{9069}$-$\forbr{Cl}{iii}{5518}{5538}} for IP$\geq$17~eV. Owing to the noisy maps of \Te(\forb{Ar}{iii}) and \Te(\forb{Ar}{iv}) (see Fig.~\ref{fig:te_ne}) we did not attempt to use them to compute ionic abundances; however, in Sect.~\ref{sec:integ_ionic_abunds} we will discuss the effect on the ionic abundances of adopting these temperatures for high ionisation potential ions (IP $>$ 35 eV).

For the \hb line, the usual methodology is to adopt \Te and \Ne selected for the specific emission line (RL or CEL) in question, following the ``classical'' approach based on eq.~3 from \citet{2023Morisset_arXiv}. 
A second order effect can be taken into account using a specific value for \Te corresponding to the {\hi} emission lines, as discussed by \citet{2023Morisset_arXiv} at the end of their Sec.~2. 
In the present case, for spatially resolved maps, this value of \Te(\hi) would fall between \Te(\forb{N}{ii}) and \Te(\forb{S}{iii}). Since these two temperatures are very close, we adhere to the classical method, where \Te(\hi) is the same as the ionic temperature of the considered ion. Regarding the integrated spectrum, we discuss the effects of using an ad-hoc value for \Te(\hi) in Sec.~\ref{sec:integ_ionic_abunds}.

\begin{figure*}
    \centering
    \includegraphics[scale = 0.47]{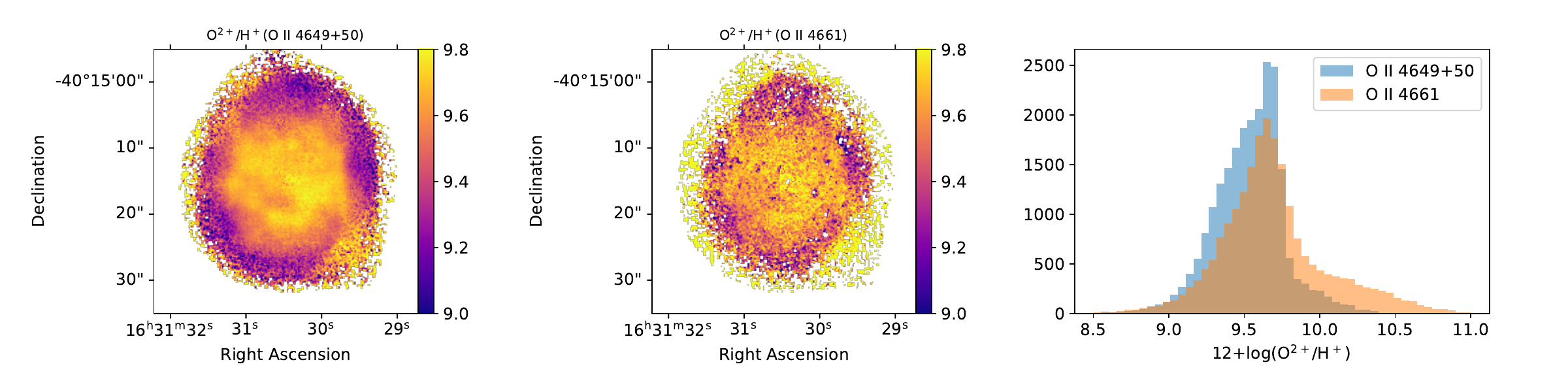}
    \caption{Histogram of the oxygen abundance derived from the \perms{O}{ii}{4649.13}{4650.85} RLs blend (blue) and from the \perml{O}{ii}{4661.63} single line (orange).}
    \label{fig:o2r_ab_hist}
\end{figure*}

The abundance map of \perms{O}{ii}{4649.13}{4650.85} is of particular interest as it is the one used to compute the ADF maps. However, considering the relatively low spectral resolution of the MUSE data and the results of \citet{liuetal00} and \citet{mcnabbetal16}, these lines can be blended with \perml{C}{iii}{4650.25}. \citet{liuetal00} reported that \perml{C}{iii}{4650.25} contributed $\sim$9 per cent to the sum of \perms{O}{ii}{4649.13}{4650.85} and \perml{C}{iii}{4650.25} integrated fluxes.
In Fig.~\ref{fig:o2r_ab_hist} we show the oxygen abundance maps derived from \perms{O}{ii}{4649.13}{4650.85} (left panel), and the lower signal-to-noise map from \perml{O}{ii}{4661.63} line (central panel).
In the right panel of Fig.~\ref{fig:o2r_ab_hist} we show the histograms of the O$^{2+}$/H$^+$ abundance ratio obtained from the \perms{O}{ii}{4649.13}{4650.85} blend (blue) and from the \perml{O}{ii}{4661.63} RL (orange). It is clear that both histograms peak at very similar values but show slightly different distributions. The long tail of \perml{O}{ii}{4661.63} corresponds to the very high values found in the most external zones of the map, where the signal-to-noise of the line is lower. However, the differences in the distributions, evident in the left wings of both histograms, are the result of different number of spaxels with detected line emissions in each scenario. In this context, no potential contamination of the \perml{C}{iii}{4650.25} line can be inferred from these histograms, given the proximity of their peaks.
 
\subsection{Weight of the cold component in the \hb emission}
\label{sec:weight}
As \citet{2015Bohigas_mnras453} and \citet{2023Morisset_arXiv} have pointed out, for PNe with two plasma components with significantly different physical conditions, the calculation of chemical abundances must take into account the weight of the line emission in both the cold and warm components, rather than assuming homogeneous physical conditions. We have followed the methodology described by \citet{2023Morisset_arXiv} to determine the chemical abundances, considering the weight of the cold region in the \hb\ emission ($\omega$, see eqs.~5, 6, 7 and 8 in \citealt{2023Morisset_arXiv}). It is worth mentioning that for the case of He ions, the case is much more complicated, since ionic He lines may be emitted by both gas phases if the cold region is not He-poor. In Sect.~\ref{sec:ionic_ab_maps_he} we discuss such a case.

For the case of NGC\,6153, the particularities of the general methodology described by \citet{2023Morisset_arXiv} are the following:

\begin{itemize}
    \item The normalised Paschen jump PJ/I$_{9229}$ is obtained for a grid of $T_{\rm e}^w$ (from 6,000 to 10,000$\,$K), $T_{\rm e}^c$ (from 1,800 to 3,000$\,$K), and $\omega$ (from 0.0 to 0.15), and constant values of $n_{\rm e}^w$ = 3,000$\,$cm$^{-3}$, $n_{\rm e}^c$ = 10$^4\,$cm$^{-3}$, He$^{+}$/H$^{+}$ = 0.1, and He$^{2+}$/H$^{+}$ = 0.005  \citep{richeretal22}.
    \item A numerical algorithm is used to interpolate the inverse problem and predict $\omega$ from: $T_{\rm e}^w$, $T_{\rm e}^c$, and PJ/I$_{9229}$.
    \item From the observed map of PJ/I$_{9229}$, $T_{\rm e}^w$ = $T_{\rm e}$(\forb{S}{iii}), and $T_{\rm e}^c$ = 2,000$\,$K, the map of $\omega$ is determined.  
\end{itemize}

The ionic abundances are then computed using eqs.~7 and 8 from \citet{2023Morisset_arXiv} :

\begin{equation}
    \left(\frac{X^i}{H^+}\right)^w = \frac{I_\lambda }{(1-\omega)\cdot I_\beta} \cdot \frac{\epsilon_\beta(T_{\rm e}^w, n_{\rm e}^w)}{\epsilon_\lambda(T_{\rm e}^w, n_{\rm e}^w)}, 
    \label{eq:ion_ab_w}
\end{equation}

for the line emitted only by the warm component (CELs) and:
\begin{equation}
    \left(\frac{X^i}{H^+}\right)^c = \frac{I_\lambda }{\omega \cdot I_\beta} \cdot \frac{\epsilon_\beta(T_{\rm e}^c, n_{\rm e}^c)}{\epsilon_\lambda(T_{\rm e}^c, n_{\rm e}^c)},
    \label{eq:ion_ab_c}
\end{equation}

for the lines emitted only by the cold component (metal RLs).

Given that $\omega$ is obtained from the comparison between \Te(PJ) and \Te(\forb{S}{iii}) \citep[see eqs.~9 and 10 from][]{2023Morisset_arXiv}), when both temperatures are close together, $\omega$ drops to very small values, leading to a very strong correction of the ionic abundance determined from the RLs due to the 1/$\omega$ factor in Eq.~\ref{eq:ion_ab_c}. These lines are actually also weakly emitted by the warm region, and $I_\lambda$ is not as small as it should be if only the contribution of the cold region were considered.  Here, we reach the limit of our fundamental assumption of the spatial origin of the emission lines, and the ionic abundances from RLs may be very overestimated. 
To avoid this effect, we cancel $\omega$ (that is, set it to {\tt Not\_a\_Number}) when \Te(PJ) reaches 85 per cent of \Te(\forb{S}{iii}), considering that in these spaxels the emission is mainly due to the warm region and that no correction can be determined.

\begin{figure*}
    \sidecaption
    \includegraphics[width=12cm]{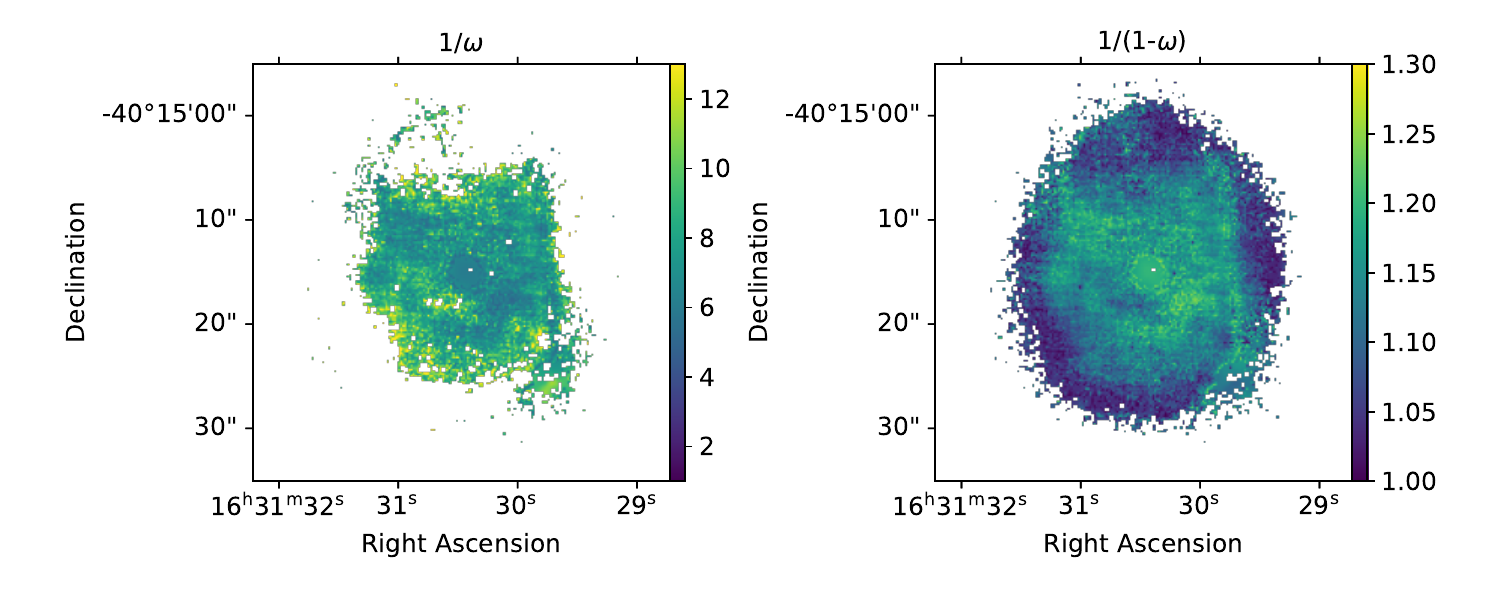}
    \caption{Corrections to apply to the ionic abundances determined the classical way, for the RLs (left panel) and the CELs (right panel), see eqs.~\ref{eq:ion_ab_w} and \ref{eq:ion_ab_c}.}
    \label{fig:omega}
\end{figure*}

In Fig.~\ref{fig:omega}, we show the spatially resolved map of 1/$\omega$ and 1/(1-$\omega$) we have obtained in NGC\,6153. 
In order to smooth the map, we have convolved it with a 2D gaussian kernel with $\sigma=2$.
These are the correction factors to apply to the abundances determined in the classical way (see eqs.~\ref{eq:ion_ab_w} and \ref{eq:ion_ab_c}). We can see that the effect of the cold region is small on the ionic abundances determined from CELs (the factor is around 1.13), while it is very important in the case of metal RLs (a factor of 10).

\begin{figure*}
    \centering
    \includegraphics[scale = 0.69]{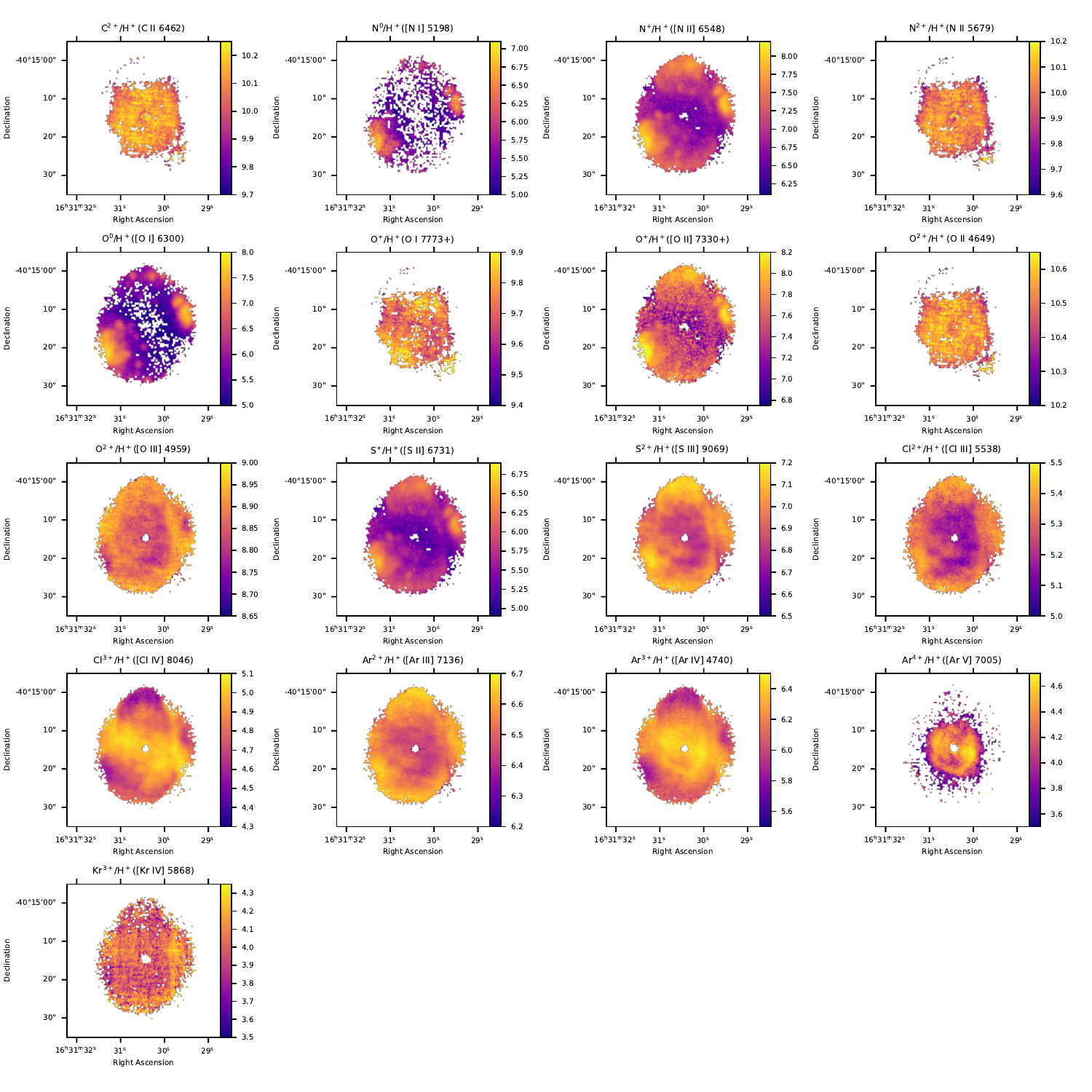}
    \caption{Ionic abundances considering $\omega$ for the \hb\ contribution. Metal RLs are considered to be totally emitted by the cold region (Eq.~\ref{eq:ion_ab_c}), while CELs are considered to be emitted by the warm region (Eq.~\ref{eq:ion_ab_w}). Helium ionic abundances cannot be determined from these equations, see Sec.~\ref{sec:ionic_ab_maps_he}.}
    \label{fig:ionic_ab_omega}
\end{figure*}

In Fig.~\ref{fig:ionic_ab_omega} we show the ionic abundance maps computed applying the parameter $\omega$ on a spaxel-by-spaxel basis. 

\begin{figure*}
    \sidecaption
    \includegraphics[width=12cm]{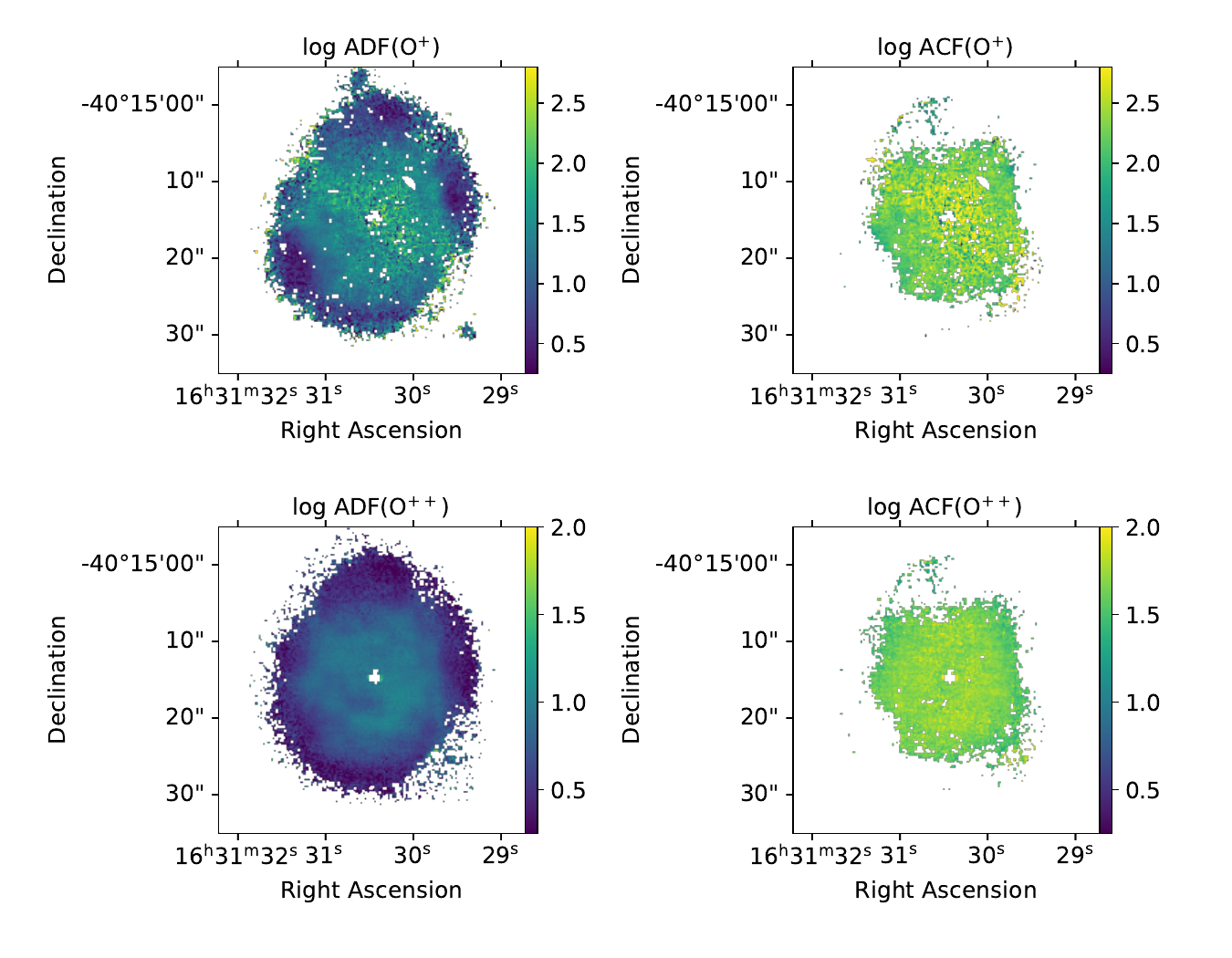}
    \caption{ADF and ACF analysis: left panels show the ADF for O$^{+}$ and O$^{2+}$ using the traditional method. Right panels display the ACF for O$^{+}$ and O$^{2+}$ considering $\omega$ for the contribution of \hb.}
    \label{fig:adf_op_opp}
\end{figure*}

Once we have the maps of $\omega$ and the ionic abundances, we generate the maps of the classical ADF, obtained from the ratio of abundance obtained from the RLs and CELs, computed using the abundance values without taking into account $\omega$. We also generate maps of the same abundance ratios, but taking into account the presence of the cold region and applying $\omega$: this is what we call the abundance contrast factor, that is, ACF \citep[see][]{2023Morisset_arXiv}. These maps are shown in Fig.~\ref{fig:adf_op_opp}, where the left (right) panels show the ADF (ACF) for O$^{+}$/H$^{+}$ (O$^{2+}$/H$^{+}$) in the upper (lower) panels. The effect of the cold region is clear in enhancing the contrast between both abundances by an average of $\simeq$ 0.9 dex. At this point, we have to emphasize that the spatial distribution of the ADF maps is equivalent to the corresponding RL/CEL line ratios. This is an important point because the different spatial distributions of the emissivities of RLs and CELs for a given ion are not evident from Fig.~\ref{fig:line_fluxes}, contrary to the objects presented in \citetalias{2022Garcia-Rojas_mnras510} (see their Figs.~S1 to S3).

\section{The integrated emission line spectrum of \ngc}
\label{sec:integrated}

To compute elemental abundances of elements heavier than helium, we have to take into account unseen ionisation stages of the different elements by adopting ionisation correction factors (ICFs). \citetalias{2022Garcia-Rojas_mnras510} pointed out that computing ICFs only makes sense when dealing with integrated spectra of the whole PN or, at least, spectra covering a significant volume of the PN that roughly includes the whole ionisation structure of the nebula. This is not the case of single spaxels of IFU observations, so elemental abundances are only computed from the integrated spectrum within the mask described in Sect.~\ref{sec:lines}. We then apply the extinction correction following the same methodology presented in Sect.~\ref{sec:extict} to obtain the extinction-corrected line fluxes. Unreddened and extinction-corrected line fluxes along with their uncertainties are presented in Table~\ref{tab:int_fluxes}. The reported extinction-corrected line fluxes of \forbl{N}{ii}{5755} and \forbl{O}{ii}{7320+30} were corrected from recombination contribution assuming \Te = 6,000\,K and \Ne = 10,000~cm$^{-3}$ (see Sect.~\ref{sec:corr_rec}). Lines affected by blends are also indicated in the table.

We used Monte Carlo simulations to estimate the uncertainties in the physical conditions and chemical abundances from the integrated spectrum. We generated 500 random values for each line intensity using a Gaussian distribution centred on the observed line intensity with a standard deviation $\sigma$ equal to its uncertainty, then the extinction coefficient, physical conditions and chemical abundances are determined for the 501 values of line intensities, and their corresponding uncertainties are determined using standard deviation or 16–84 percentiles.

\subsection{Physical conditions}
\label{sec:int_phys_cond}

We computed the physical conditions (\Te and \Ne) from the integrated spectrum using several pairs of diagnostics. As mentioned in Sect.~\ref{sec:corr_rec}, and following the same procedure, we have considered a 40 per cent recombination contribution to the auroral \forbl{N}{ii}{5755} line for the computation of \Te({\forb{N}{ii}}), leading to a corrected \Te\ of 8760\,K, i.e, $\sim$1800\,K lower than without taking into account such a correction. Similarly, we also applied this procedure to the intensities of the \forbl{O}{ii}{7320+30} which leads to a 51 per cent recombination contribution. In Table~\ref{tab:int_fluxes}, we present for these lines both the unreddened measured fluxes and the intensities corrected for extinction and recombination contribution.

As is well known, the main drawback of MUSE spectra of photoionised nebulae is the lack of coverage of wavelengths bluer than 4600 \AA\, preventing the detection of the auroral [O~{\sc iii}] $\lambda$4363 line, which is essential for computing the widely used \Te([O~{\sc iii}]). However, the depth of the MUSE NGC\,6153 spectrum allows the detection of additional \Te-sensitive auroral lines in the integrated spectrum with enough signal-to-noise to compute electron temperatures. This is exemplified by the faint [Ar~{\sc iii}] $\lambda$5192 and [Ar~{\sc iv}] $\lambda$7170 lines. The calculated physical conditions are presented in Table~\ref{tab:int_tene}. To compute \Te(PJ) and \Te({\hei}) we assumed the density provided by the (\Te([S~{\sc iii}]), \Ne([S~{\sc ii}]) pair. We also computed an averaged value of \Te([Ar~{\sc iii}]) and \Te([Ar~{\sc iv}]), given by:

\begin{equation}\label{eq:ar_average}
T_{\rm e}({\rm Ar}) = \theta \cdot T_{\rm e}([{\rm Ar~III}]) + (1-\theta)\cdot T_{\rm e}([{\rm Ar~IV}]), 
\end{equation}
where $\theta$ is the abundance ratio Ar$^{2+}$/(Ar$^{2+}$+Ar$^{3+}$). Given the ionisation potential (IP) ranges of Ar$^{2+}$ (27.63--40.79 eV) and Ar$^{3+}$ (40.79--59.58 eV) and the IP range of O$^{2+}$ (35.12--54.94 eV), this \Te diagnostic could better mimic the physical conditions in the ionisation range covered by the missing \Te([O~{\sc iii}]) than any of the available \Te diagnostics; for example, \Te([S~{\sc iii}]) would be reliable for ions covering an IP range between 23.34--34.79 eV) and, therefore would be only representative of the mid-ionisation region in the nebula. 

The electron temperature derived from the [S~{\sc iii}] line ratio is lower than that obtained from other ions. We have used the PyNeb default set of atomic data labelled \texttt{PYNEB\_23\_01}. We made the test of modifying the default [S~{\sc iii}] effective collision strengths from \citet{tayalgupta99} to \citet{2014Grieve_apj780}, and obtained a \Te([S~{\sc iii}]) $\simeq$ 300 K higher, which would be in better agreement with what is obtained from the [Ar~{\sc iii}] lines. The effect of these new atomic data in the ionic abundance maps shown in Fig.~\ref{fig:ionic_ab_omega} would be relatively small, and the resulting spatial distributions would be practically unaffected. 

On the other hand, since we lack the detection of auroral [O~{\sc iii}] lines, we rely on the average of [Ar~{\sc iii}] and [Ar~{\sc iv}] temperatures, which is quite consistent with the values of \Te([O~{\sc iii}]) derived by \citet[][see left panel of their Fig. 37]{richeretal22}. The differences between \Te([O~{\sc iii}]) and \Te([Ar~{\sc iii}]), which is also consistent with our derived \Te([Ar~{\sc iii}]), are attributed by these authors to the effect of temperature fluctuations in the warm component.

\begin{table*}
    \centering
    \caption{Physical conditions in the integrated spectra.}
    \label{tab:int_tene}
    \begin{tabular}{lcc}
        \hline
        Diagnostic   & $T_{\rm e}$ [K] & $n_{\rm e}$ [cm$^{-3}$] \\
        \hline
\Te(\forb{N}{ii} 5755/6584),  \Ne(\forb{S}{ii} 6716/6731)     &  8760 $\pm$  350 & 4210 $\pm$ 1780 \\
\Te(\forb{S}{iii} 6312/9069),  \Ne(\forb{S}{ii} 6716/6731)    &  7970 $\pm$  230 & 3910 $\pm$ 1780 \\
\Te(\forb{S}{iii} 6312/9069),  \Ne(\forb{Cl}{iii} 5518/5538)  &  8060 $\pm$  220 & 3410 $\pm$  590 \\
\Te(\forb{S}{iii} 6312/9069),  \Ne(\forb{Ar}{iv} 4711/4740)   &  7930 $\pm$  230 & 7790 $\pm$ 1490 \\
\Te(\forb{Ar}{iii} 5192/7136),  \Ne(\forb{S}{ii} 6716/6731)   &  8670 $\pm$  200 & 4120 $\pm$ 1700 \\
\Te(\forb{Ar}{iii} 5192/7136),  \Ne(\forb{Cl}{iii} 5518/5538) &  8610 $\pm$  195 & 3500 $\pm$  610 \\
\Te(\forb{Ar}{iv} 4740/7170),  \Ne(\forb{S}{ii} 6716/6731)    & 12490 $\pm$  670 & 4800 $\pm$ 1820 \\
\Te(\forb{Ar}{iv} 4740/7170),  \Ne(\forb{Cl}{iii} 5518/5538)  & 12390 $\pm$  700 & 3810 $\pm$  700 \\
\Te(\forb{Ar}{iv} 4740/7170),  \Ne(\forb{Ar}{iv} 4711/4740)   & 12530 $\pm$  690 & 9200 $\pm$ 1760 \\
\Te(average \forb{Ar}{iii}, \forb{Ar}{iv})                    &  9590 $\pm$  195 & --- \\
PJ                                                            &  6180 $\pm$ 1250 & --- \\
{\hei} $\lambda$7281/$\lambda$6678                            &  5160 $\pm$  570 & --- \\
\hline
    \end{tabular}

\end{table*}

\subsection{Ionic abundances}
\label{sec:integ_ionic_abunds}
To compute the ionic abundances from the integrated spectra, we select the same set of emission lines as in the case of ionic abundance maps (see Sect.~\ref{sec:ionic_ab_maps}). We explore the effect of considering the weight of the cold component in the \hb emission in the chemical abundances by applying eqs.~\ref{eq:ion_ab_w} and~\ref{eq:ion_ab_c} for CELs-based and ORLs-based abundances, respectively. The contribution of recombination to the intensities of {\forbl{O}{ii}{7320+30}} were considered following the same methodology as in Sect.~\ref{sec:corr_rec}.

We explore 7 different \Te-\Ne recipes to compute ionic abundances. The first 3 recipes do not take into account $\omega$, while recipes 4 to 7 use it. The description of all the recipes is available in Sect.~\ref{sec:recipes} of the appendix. The ionic abundances obtained with these 7 recipes are given in Tab.~\ref{tab:recipes} (as well as the temperatures adopted in each recipe).

\begin{table*}
    \small
    \centering
    \caption{Ionic abundances (in units of 12 + log(X$^i$/H$^+$)) obtained using the different recipes described in the text$^{\rm a}$. }
    \begin{tabular}{llccccccc} 
        \hline
         & & \multicolumn{7}{c}{Recipe} \\
         & & 1 & 2 & 3 & 4 & 5 &  6&7\\ 
        \hline
         \multicolumn{2}{l}{Use $\omega$} &  no&  no&  no&  yes&  yes&  yes&yes\\ 
        \hline
         \multicolumn{2}{l}{\Te(RLs)$^{\rm b}$}&  \forb{N}{ii}/\forb{S}{iii}&  2000&  2000&  2000&  2000&  2000&2000\\ 
         \multicolumn{2}{l}{\Te(\hi\xspace in RLs)$^{\rm c}$} & Ion&  Ion&  PJ&  Ion&  Ion&  Ion&Ion\\ 
        \hline
         \multicolumn{2}{l}{\Te(CELs)$^{\rm b}$}&  \forb{N}{ii}/\forb{S}{iii}&   \forb{N}{ii}/\forb{S}{iii}&  \forb{N}{ii}/\forb{S}{iii}&  \forb{N}{ii}/\forb{S}{iii}&  \forb{N}{ii}/\forb{S}{iii}&  \forb{N}{ii}/\forb{S}{iii}&\forb{N}{ii}/\forb{S}{iii}\\ 
         \multicolumn{2}{l}{\Te(\hi\xspace in CELs)$^{\rm c}$} & Ion&  Ion& PJ&  Ion&  Ion&  Ion &8300\\ 
         \multicolumn{2}{l}{\Te(CELs IP > 35 eV)} &  \forb{S}{iii}& \forb{S}{iii}&  \forb{S}{iii}& \forb{S}{iii}& \forb{Ar}{iii}& <\forb{Ar}{iii}, \forb{Ar}{iv}>&<\forb{Ar}{iii}, \forb{Ar}{iv}>\\ 
         \hline
         Ab. ratio& Line & \multicolumn{7}{c}{12+log(X${i+}$/H$^+$) from CELs}  \\
\hline
N$^{0}$/H$^+$ & \forb{N}{i} 5198& 5.79$\pm$0.12& 5.79$\pm$0.12& 5.92$\pm$0.12& 5.84$\pm$0.12& 5.84$\pm$0.12& 5.84$\pm$0.12& 5.86$\pm$0.10 \\
N$^{+}$/H$^+$ & \forb{N}{ii} 6548& 7.13$\pm$0.07& 7.13$\pm$0.07& 7.26$\pm$0.09& 7.18$\pm$0.07& 7.18$\pm$0.07& 7.18$\pm$0.07& 7.20$\pm$0.05 \\
O$^{0}$/H$^+$ & \forb{O}{i} 6300& 6.42$\pm$0.08& 6.42$\pm$0.08& 6.55$\pm$0.09& 6.47$\pm$0.08& 6.47$\pm$0.08& 6.47$\pm$0.08& 6.49$\pm$0.06 \\
O$^{+}$/H$^+$ & \forb{O}{ii} 7330+& 7.54$\pm$0.13& 7.54$\pm$0.13& 7.67$\pm$0.14& 7.59$\pm$0.13& 7.59$\pm$0.13& 7.59$\pm$0.13& 7.61$\pm$0.12 \\
O$^{2+}$/H$^+$ & \forb{O}{iii} 4959& 8.83$\pm$0.05& 8.83$\pm$0.05& 8.93$\pm$0.08& 8.88$\pm$0.05& 8.77$\pm$0.04& 8.59$\pm$0.04& 8.65$\pm$0.04 \\
S$^{+}$/H$^+$ & \forb{S}{ii} 6731& 5.71$\pm$0.09& 5.71$\pm$0.09& 5.84$\pm$0.10& 5.76$\pm$0.09& 5.76$\pm$0.09& 5.76$\pm$0.09& 5.78$\pm$0.08 \\
S$^{2+}$/H$^+$ & \forb{S}{iii} 9069& 6.91$\pm$0.05& 6.91$\pm$0.05& 7.01$\pm$0.08& 6.96$\pm$0.05& 6.96$\pm$0.05& 6.96$\pm$0.05& 6.95$\pm$0.04 \\
Cl$^{2+}$/H$^+$ & \forb{Cl}{iii} 5538& 5.23$\pm$0.05& 5.23$\pm$0.05& 5.32$\pm$0.08& 5.28$\pm$0.05& 5.28$\pm$0.05& 5.28$\pm$0.05& 5.27$\pm$0.04 \\
Cl$^{3+}$/H$^+$ & \forb{Cl}{iv} 8046& 4.85$\pm$0.04& 4.85$\pm$0.04& 4.94$\pm$0.08& 4.90$\pm$0.04& 4.82$\pm$0.04& 4.71$\pm$0.03& 4.76$\pm$0.03 \\
Ar$^{2+}$/H$^+$ & \forb{Ar}{iii} 7136& 6.49$\pm$0.04& 6.49$\pm$0.04& 6.58$\pm$0.08& 6.54$\pm$0.04& 6.54$\pm$0.04& 6.54$\pm$0.04& 6.53$\pm$0.04 \\
Ar$^{3+}$/H$^+$ &\forb{Ar}{iv} 4740& 6.21$\pm$0.05& 6.21$\pm$0.05& 6.31$\pm$0.08& 6.26$\pm$0.05& 6.14$\pm$0.04& 5.97$\pm$0.05& 6.02$\pm$0.05 \\
Ar$^{4+}$/H$^+$ & \forb{Ar}{v} 7005& 3.89$\pm$0.05& 3.89$\pm$0.05& 3.98$\pm$0.08& 3.94$\pm$0.05& 3.85$\pm$0.04& 3.72$\pm$0.03& 3.77$\pm$0.03 \\
Kr$^{3+}$/H$^+$ & \forb{Kr}{iv} 5868& 3.69$\pm$0.05& 3.69$\pm$0.05& 3.79$\pm$0.08& 3.74$\pm$0.05& 3.65$\pm$0.04& 3.50$\pm$0.04& 3.56$\pm$0.03 \\
\hline
         Ab. ratio& Line & \multicolumn{7}{c}{12+log(X${i+}$/H$^+$) from RLs} \\
\hline
He$^+$/H$^+$ & \perm{He}{i} 5876 &11.08$\pm$0.03&11.08$\pm$0.03&11.02$\pm$0.09& \multicolumn{4}{c}{ 10.95 (w),  11.48 (c)$^{\rm d}$}  \\
He$^{2+}$/H$^+$ & \perm{He}{ii} 4686 &10.03$\pm$0.03&10.03$\pm$0.03&10.13$\pm$0.08& \multicolumn{4}{c}{ $<$ 10.08 (w), $<$ 10.86 (c)$^{\rm d}$} \\
C$^{2+}$/H$^+$ & \perm{C}{ii} 6462& 9.31$\pm$0.02& 9.16$\pm$0.02& 8.79$\pm$0.07&10.13$\pm$0.02&10.13$\pm$0.02&10.13$\pm$0.02&10.13$\pm$0.02 \\
N$^{2+}$/H$^+$ & \perm{N}{ii} 5679& 9.13$\pm$0.02& 9.07$\pm$0.02& 8.70$\pm$0.07&10.04$\pm$0.02&10.04$\pm$0.02&10.04$\pm$0.02&10.04$\pm$0.02 \\
O$^{+}$/H$^+$ & \perm{O}{i} 7773+& 8.82$\pm$0.04& 8.78$\pm$0.04& 8.41$\pm$0.08& 9.75$\pm$0.04& 9.75$\pm$0.04& 9.75$\pm$0.04& 9.75$\pm$0.04 \\
O$^{2+}$/H$^+$ & \perm{O}{ii} 4649+& 9.65$\pm$0.03& 9.59$\pm$0.03& 9.21$\pm$0.08&10.56$\pm$0.03&10.56$\pm$0.03&10.56$\pm$0.03&10.56$\pm$0.03 \\
O$^{2+}$/H$^+$ & \perm{O}{ii} 4661& 9.52$\pm$0.03& 9.57$\pm$0.03& 9.19$\pm$0.08&10.53$\pm$0.03&10.53$\pm$0.03&10.53$\pm$0.03&10.53$\pm$0.03 \\

\hline
\end{tabular}
\begin{description}
\item $^{\rm a}$ {The first 6 rows summarise the temperatures used in each recipe. The following rows give the ionic abundance obtained from the emission line given in the corresponding first column.} 
\item $^{\rm b}$ {$T_{\rm e}$(\forb{N}{ii}) for ions with IP < 17 eV / $T_{\rm e}$(\forb{S}{iii}) for ions with IP > 17 eV.}
\item $^{\rm c}$ {``Ion'' meaning that for computing the emissivity of \hi\ we adopt the same physical conditions than for the corresponding ion; ``PJ'' is the $T_{\rm e}$ computed from the Paschen jump.}
\item $^{\rm d}$ {For the warm (w) and cold (c) regions. See Sect.~\ref{sec:helium_integ}.}
\end{description}
\label{tab:recipes}
\end{table*}

The differences between recipes 1 to 3 in the abundances are owing to the different electron temperatures being used on each recipe. Recipes 1 and 2 are the same for the CELs; in the case of metal RLs, the small difference in ionic abundances comes from the relative differences between the corresponding metal line emissivity and the \hb emissivity, when changing the temperatures from \Te({\forb{N}{ii}}, {\forb{S}{iii}}) to 2,000~K.
We find a larger difference for recipe 3, where \Te(PJ) was adopted to compute the \hb emissivity.  Compared to recipe 2, the abundances from CELs increase by $\simeq 0.1$ dex, while the abundances determined from RLs decrease by $\simeq 0.35$ dex. Although recipe 3 might seem reasonable, in the case of two plasma components this temperature is not representative of the emissivity of \hi\xspace in either component.

The difference between recipes 1 to 3 and 4 to 7 is mainly due to the use of Eqs.~\ref{eq:ion_ab_w} and \ref{eq:ion_ab_c} in the latter recipes, especially the use of $\omega$: RL-based ionic abundances are enhanced by $\simeq$ 0.9 dex, while the CEL-based ionic abundances are only enhanced by $\simeq$ 0.05 dex.

In the case of recipes 4 to 7, RL abundances are always computed using the same temperature corresponding to the cold region. For the CEL abundances, the differences are only important for the highest excitation ions (with IP > 35 eV), as the use of higher \Te than that provided by \Te({\forb{S}{iii}}), such as \Te(\forb{Ar}{iii}) or the average of \Te(\forb{Ar}{iii}) and  \Te(\forb{Ar}{iv}) translates into lower ionic abundances, by $\simeq$ 0.1 dex or $\simeq$ 0.25 dex, respectively.

Our favoured recipe is the number 7, where each ion emissivity (including H$^+$) is computed using its corresponding physical conditions, and the factor $\omega$ is taken into account.

\subsection{The helium case}
\label{sec:helium_integ}

The case of the helium ionic abundances is described in Sect.~\ref{sec:ionic_ab_maps_he}. We apply here the same equations \ref{eq:he1warm} and \ref{eq:he1cold} to determine (He$^+$/H$^+$)$^w$ = 0.0892 and (He$^+$/H$^+$)$^c$= 0.303 for the integrated spectrum.

In the case of \heii we have only one line, and there is no unique solution to the single equation \ref{eq:ion_ab_2comp}.
The extreme values can be determined: 0 < (He$^{2+}$/H$^+$)$^w$ < 0.0120, and 0 < (He$^{2+}$/H$^+$)$^c$ < 0.073. 
Given that He$^{2+}$ is a residual ion, the exact value of its abundance does not affect the elemental He / H ratio too much. It is nevertheless an ingredient in the determination of the ICFs used to compute the elemental abundances.

The determination of He/H for both regions is then: 0.089 < (He/H)$^w$ < 0.10, and 0.30 < (He/H)$^c$ < 0.38.
The overabundance of helium in the cold region is much smaller than that of the metals. This result confirms the upper limit of He/H < 6 in the cold region, obtained by \citet{2020Gomez-Llanos_mnra497}.
The emission of the dominant ion He$^+$ that comes from each region are comparable: 60 per cent (70 per cent) of the \hei 6678 (7281, respectively) comes from the warm phase. 

\subsection{Total abundances}
\label{sec:integ_total_abunds}

Following our discussion in the previous section, we consider computing the total ionic abundances using a single method, taking into account $\omega$, specifically recipe 7. This approach involves three ionisation zones, a characteristic \Te for hydrogen in the warm phase, and a low \Te in the cold phase, to best represent the actual values of ionic abundances.

To calculate the elemental abundances of elements heavier than helium in the warm phase, an ICF should be computed to account for unobserved ionisation states. Similarly to \citetalias{2022Garcia-Rojas_mnras510}, in this work we have taken three different approaches to calculate ICFs: i) using ICFs from the literature; ii) looking for ICFs from photoionisation models close to the observed object, and iii) computing ad hoc ICFs using machine learning techniques.

In the first approach, for N, O, S, Cl, and Ar we adopt the ICFs provided by \citet{delgadoingladaetal14} from their equations 14, 12, 26, 29, and 35/36, respectively. Uncertainties were computed by using the analytic expressions recommended by \citet{delgadoingladaetal14} for each case, which are based on the maximum dispersion of each ICF obtained from their grid of photoionisation models. \citet{delgadoingladaetal15} highlighted that the calculated uncertainties in the overall abundances, based on the analytical expressions from \citet{delgadoingladaetal14}, are likely overestimated. This is because the maximum dispersion within each ICF is employed, and these ICFs have been derived from a broad grid of photoionisation models. However, real nebulae are probably more accurately depicted by a smaller selection of models.
For Kr we use equation 3 by \citet{sterlingetal15}. To compute the final uncertainties associated with these ICFs, we consider both the uncertainties associated with the physical conditions and the ionic abundances and those related to the adopted ICF. We follow the same methodology as in \citet{delgadoingladaetal15} and generate a uniform distribution of the ICFs with maximum and minimum values given by the uncertainties computed in Table~\ref{tab:icfs}, and then generate random values that we use to calculate the total abundances for 500 Monte Carlo  simulations, adopting as the final errors for each abundance the 16 and 84 percentiles. 

In the second method (look-up table method), we use the \verb!3MdB_17! database \citep{morissetetal15} to search for models that are representative of the ionisation state of NGC~6153: we select the models that are between the 5 and 95 percentiles of the following ionic abundance ratios: O$^{2+}$/O$^{+}$, S$^{2+}$/S$^{+}$, Cl$^{3+}$/Cl$^{2+}$ and Ar$^{3+}$/Ar$^{2+}$, determined from the observations using recipe 7 and 500 Monte Carlo simulations (see Sect.~\ref{sec:integ_ionic_abunds}, and Tab.~\ref{tab:recipes}). 
We only search within a subset of models from \verb!3MdB_17! under the references \verb!ref like "PNe_202_"!, with \verb!com6 = 1! \citep[realistic models, see][]{delgadoingladaetal14}\footnote{And also \href{https://sites.google.com/site/mexicanmillionmodels/the-different-projects/pne_2014}{https://sites.google.com/site/mexicanmillionmodels/the-different-projects/pne\_2014}} 
and \verb!MassFrac > 0.7! (radiation- and matter-bounded models with at least 70 per cent of the mass of the corresponding radiation-bounded model).
These criteria lead to the selection of 247 models from which we determine the median, the 16 per cent, and the 84 per cent quantiles of the distribution of the needed ICFs.

To estimate the ad hoc ICFs from the third method, we follow the methodology described in Sect. 6.5.1 of \citetalias{2022Garcia-Rojas_mnras510}. We select a subset of photoionisation models from the \verb!3MdB_17! database \citep{morissetetal15} to train an Artificial Neural Network (ANN) to predict the needed ICFs from some ionic abundance ratios. In this case, we select models that fit the observed values of the O$^{2+}$/O$^{+}$, S$^{2+}$/S$^{+}$, Cl$^{3+}$/Cl$^{2+}$ and Ar$^{3+}$/Ar$^{2+}$ ionic abundance ratios within 0.5 dex, and He$^{2+}$/He$^{+}$ abundance ratio within 0.1 dex. We use 80 per cent of this set of models to train the ANN (16,756 models) and test the prediction of the ANN on the remaining 20 per cent (4,189 models). The input vector is built from the five ionic abundance ratios described above. The output vector contains eight ICFs: ICF(N$^+$/H$^+$), ICF((O$^+$ + O$^{2+}$)/H$^+$), ICF((S$^+$ + S$^{2+}$)/H$^+$), ICF((Cl$^{2+}$ + Cl$^{3+}$)/H$^+$), ICF((Ar$^{2+}$ + Ar$^{3+}$)/H$^+$), ICF(N$^+$/O$^+$), ICF((S$^+$ + S$^{2+}$)/O$^+$), and ICF(Ar$^{2+}$/(O$^+$ + O$^{2+}$)). The ANN is obtained using the SciKit-Learn Python library. It contains 3 dense layers of 50, 80, and 80 RELU perceptrons, respectively. Convergence is reached using the ADAM algorithm.
We consider the He$^{+}$/H$^+$ abundance for the warm region computed from eq.~\ref{eq:he1warm} with 500 Monte Carlo simulations, and set the He$^{2+}$/H$^+$ ionic abundance as a random uniform distribution between 0 and 0.012. The exact value does not strongly affect the results. To estimate the uncertainties with this method, we compute the ICFs using the observed values of the input ionic fraction and 500 Monte Carlo simulations for He$^{2+}$/He$^{+}$, O$^{2+}$/O$^{+}$, S$^{2+}$/S$^{+}$, Cl$^{3+}$/Cl$^{2+}$ and Ar$^{3+}$/Ar$^{2+}$, respectively. We then determine the 16 per cent and 84 per cent quantiles of the distribution of the ICFs computed by the ANN.

\begin{figure*}
    \centering
    \includegraphics[width=5.5cm]{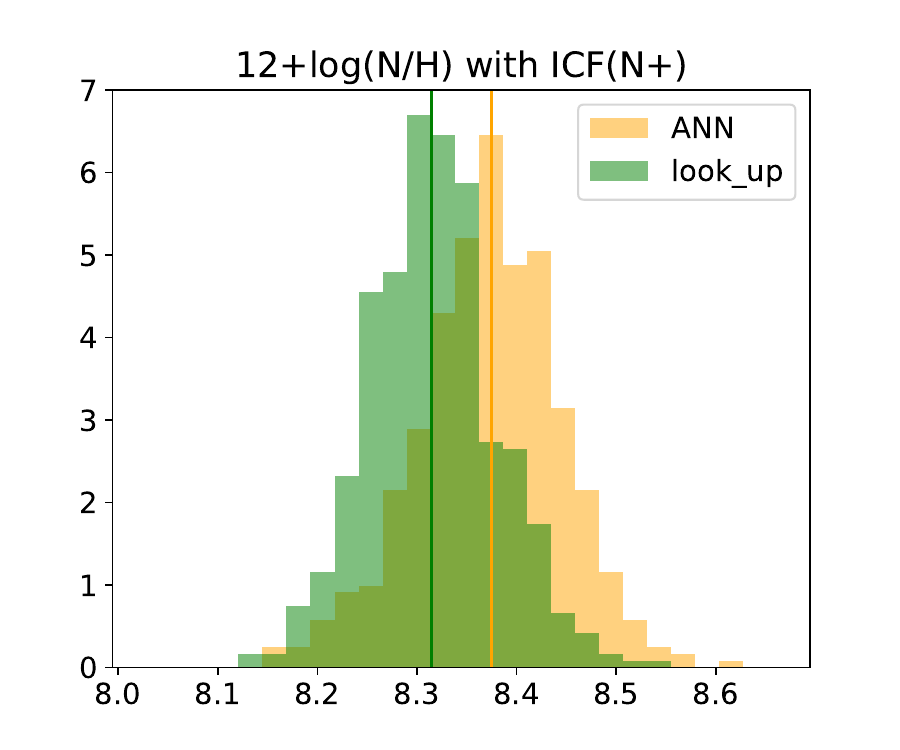}
    \includegraphics[width=5.5cm]{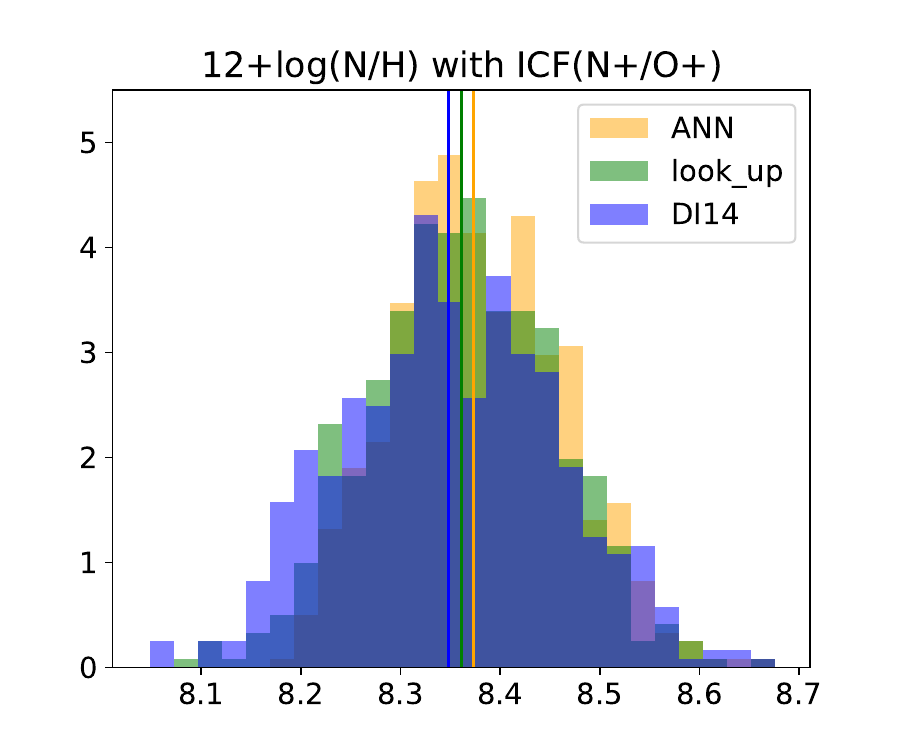}
    \includegraphics[width=5.5cm]{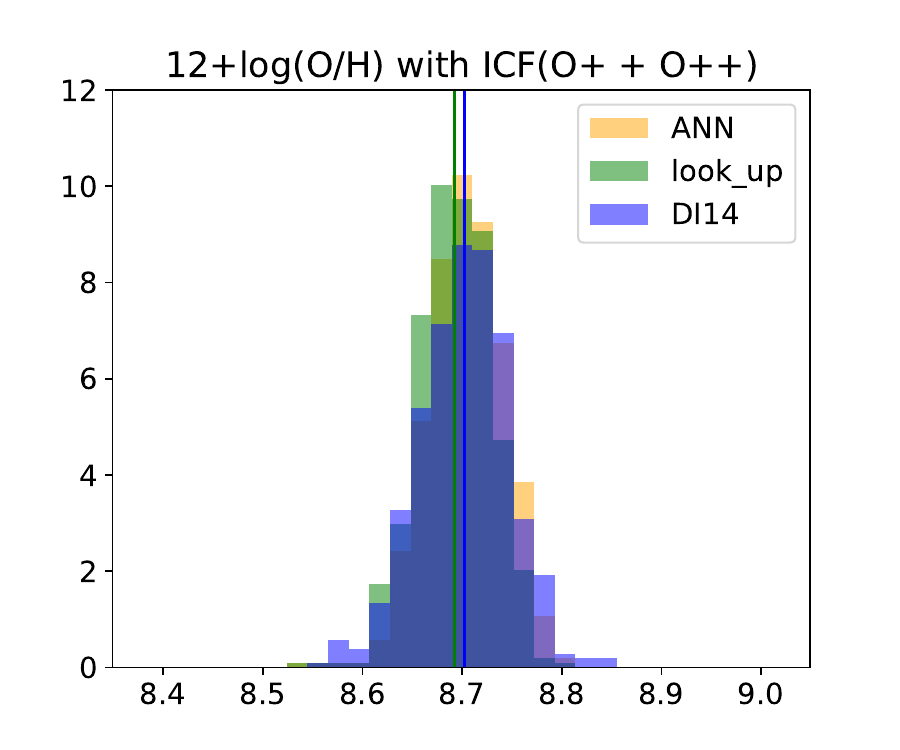}
    \includegraphics[width=5.5cm]{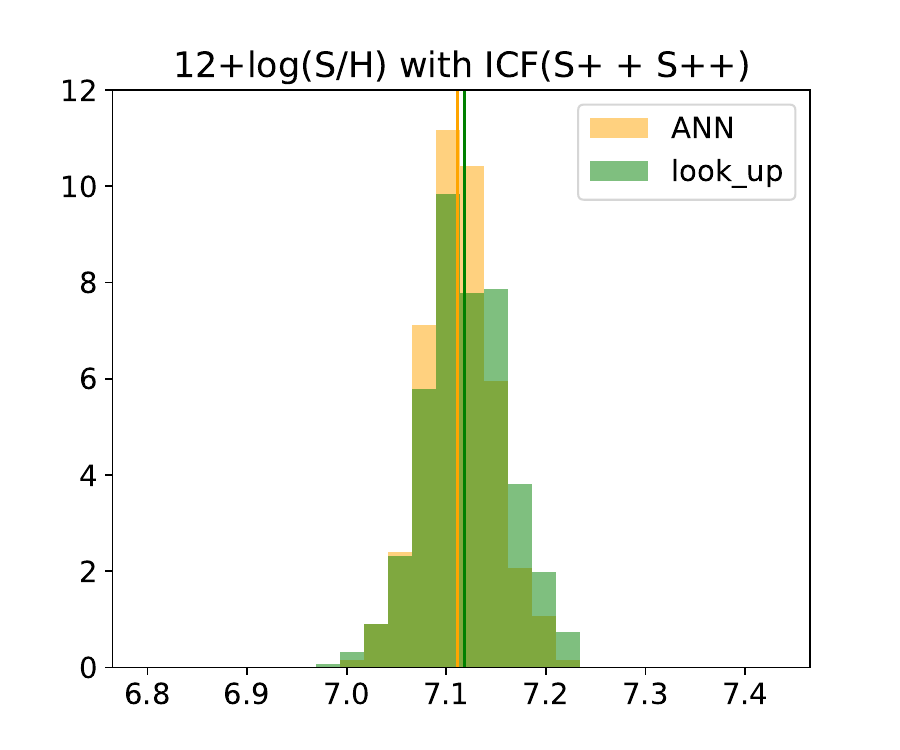}
    \includegraphics[width=5.5cm]{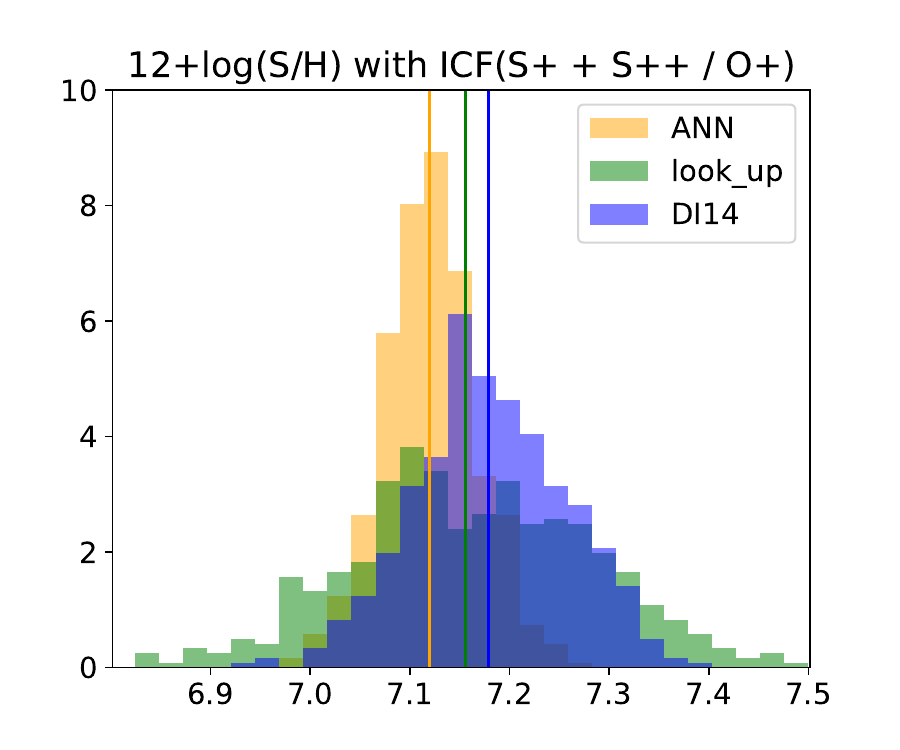}
    \includegraphics[width=5.5cm]{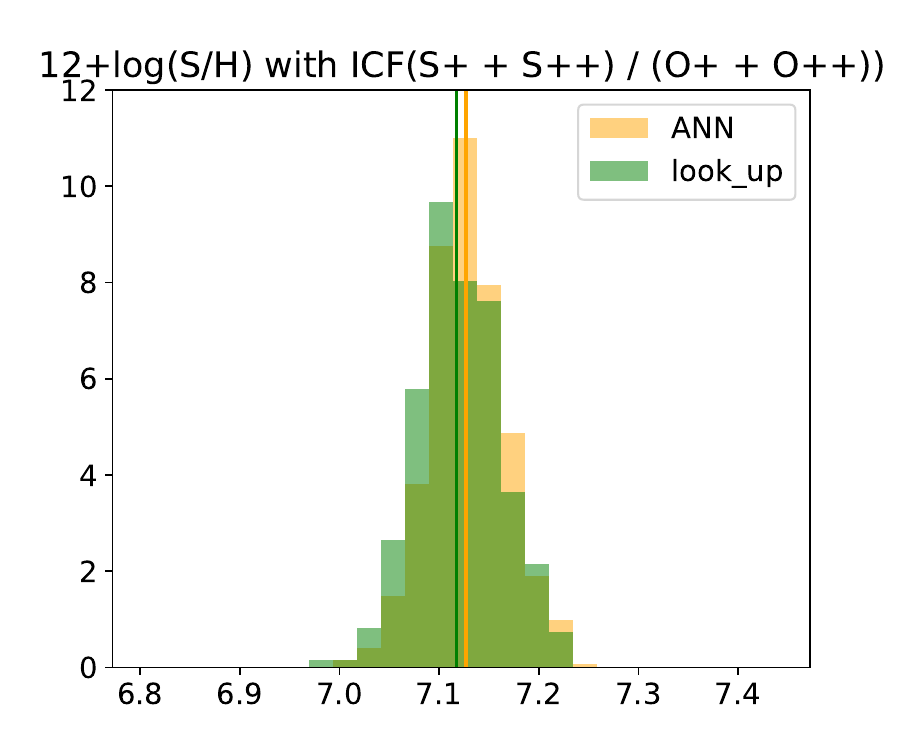}
    \includegraphics[width=5.5cm]{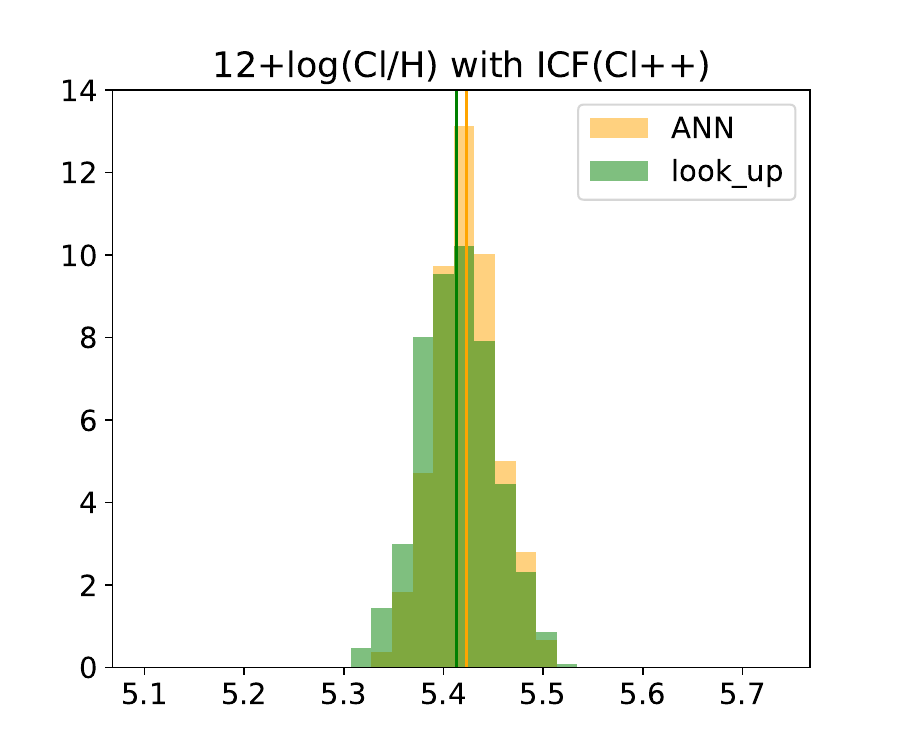}
    \includegraphics[width=5.5cm]{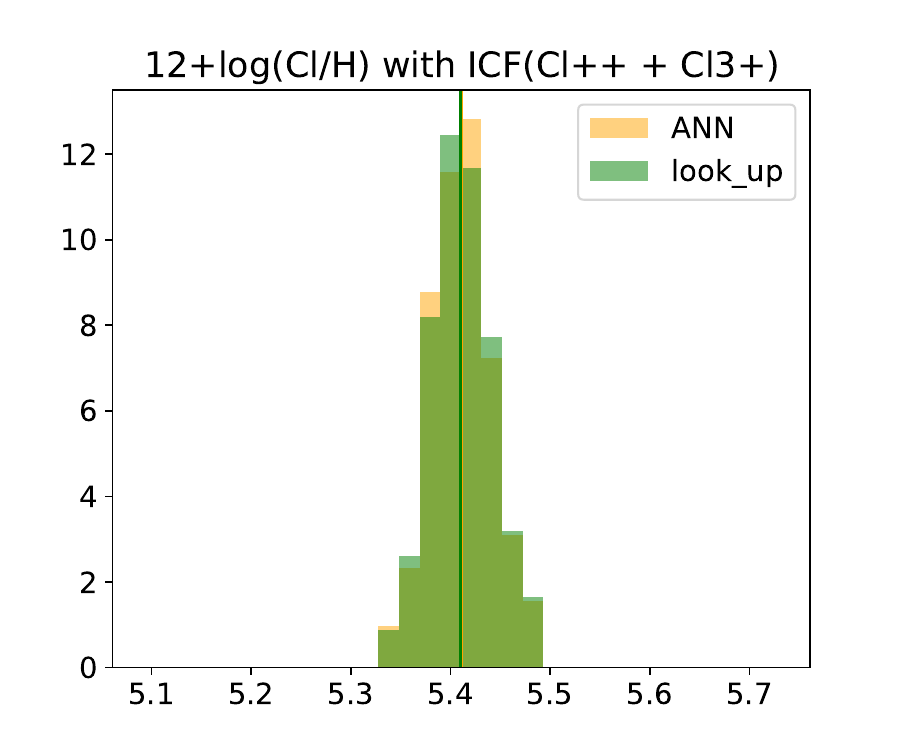}
    \includegraphics[width=5.5cm]{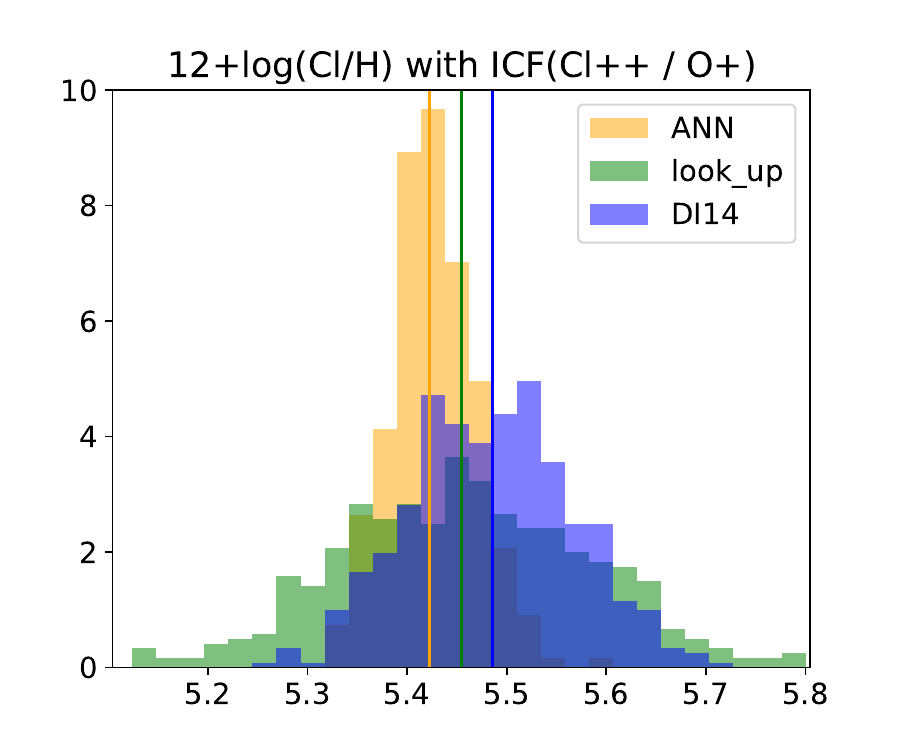}
    \includegraphics[width=5.5cm]{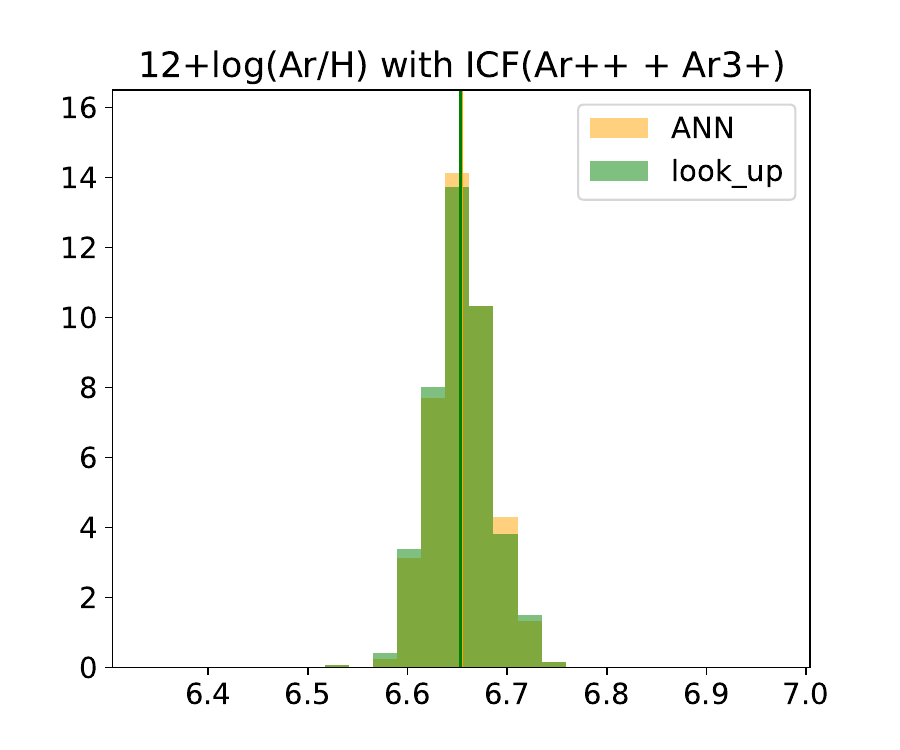}
    \includegraphics[width=5.5cm]{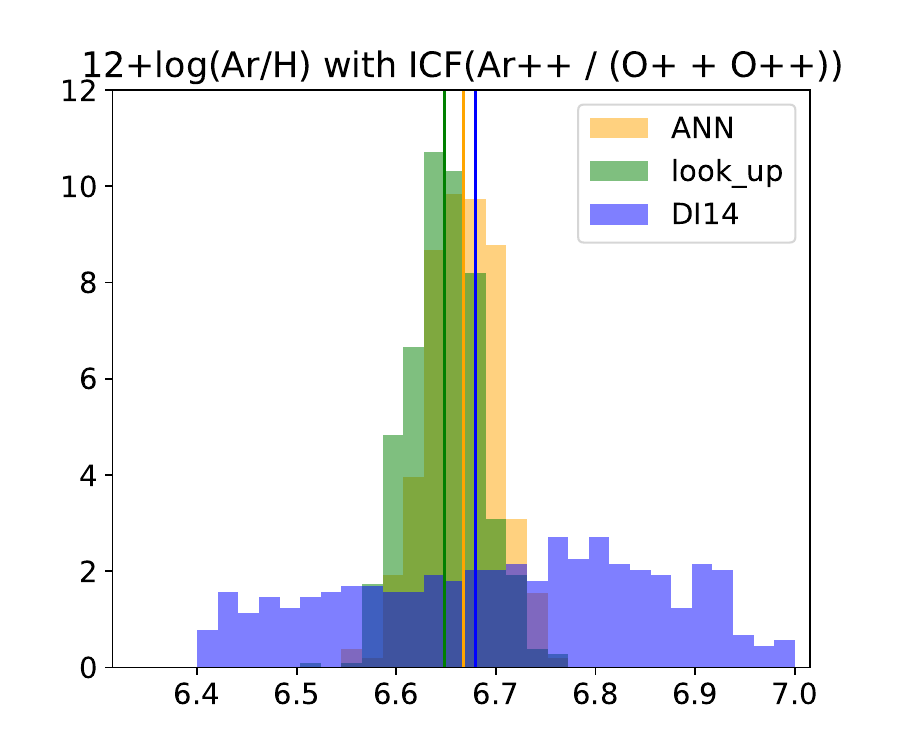}
    \caption{Monte Carlo distributions of the total abundances derived from CELs with the three methods for the ICFs: literature ICFs from \citet{delgadoingladaetal14} (purple), look-up method (green) and ANN (orange). See Table~\ref{tab:icfs} for the labels of the ICFs. Vertical lines show the median of each distribution. Only ICFs obtained with more than one method are shown.}
    \label{fig:hist_abun}
\end{figure*}

For the cold gas component, we used only the ICFs from \citet{delgadoingladaetal14} for C (equation 39) and O (equation 12), given the lack of additional ionisation-stage diagnostics on this plasma component.   
In Tab.~\ref{tab:icfs} we summarise the ICFs we obtain using the 3 methods. We do not find any systematic differences between the three methods, which provide values that are very coherent with each other. 

\begin{table*}
    \centering
    \caption{Ionisation correction factors (ICFs) obtained using three different methodologies (see text).}
    \label{tab:icfs}
    \begin{tabular}{llccc}
        \hline
        Ab ratio & ICF & literature & look-up table & ANN \\
        \hline
        \\[-1em]
        N/H & N$^+$/H$^+$                           & ---      & $12.72^{+ 1.43}_{-1.17}$ & $14.69^{+ 3.06}_{- 2.97}$\\
        \\[-.5em]
        N/O & N$^+$/O$^+$                           & $1.16^{+ 0.21}_{-0.12}$ &$ 1.24^{+ 0.08}_{-0.07}$ & $ 1.27^{+ 0.13}_{- 0.12}$ \\
        \\[-.5em]
        O/H & (O$^+$ + O$^{2+}$)/H$^+$              & $1.03^{+ 0.07}_{-0.06}$ & $ 1.01\pm0.01$ & $ 1.04\pm0.02$  \\
        \\[-.5em]
        S/H & (S$^+$ + S$^{2+}$)/H$^+$              & ---      &$ 1.34^{+ 0.04}_{-0.03}$ & $ 1.31^{+ 0.06}_{- 0.04}$  \\
        \\[-.5em]
        S/O & (S$^+$ + S$^{2+}$)/O$^+$              & $0.13^{+ 0.03}_{-0.02}$     & $ 0.13^{+ 0.02}_{-0.01}$ & $ 0.11^{+ 0.03}_{- 0.02}$ \\
        \\[-.5em]
        S/O & (S$^+$ + S$^{2+}$)/(O$^+$ + O$^{2+}$) & ---      &$ 1.33\pm0.04$ & $ 1.32^{+ 0.07}_{- 0.06}$  \\
        \\[-.5em]
        Cl/H & Cl$^{2+}$/H$^+$                      & ---      &$ 1.38^{+ 0.03}_{-0.02}$ & $ 1.41^{+ 0.05}_{- 0.04}$\\
        \\[-.5em]
        Cl/H & (Cl$^{2+}$ + Cl$^{3+}$)/H$^+$        & ---      &$ 1.05\pm0.01$ & $ 1.05\pm 0.01$ \\
        \\[-.5em]
        Cl/O & Cl$^{2+}$/O$^+$                      & $0.14^{+ 0.04}_{-0.03}$ $^a$ & $ 0.13^{+ 0.02}_{-0.01}$ & $ 0.12^{+ 0.03}_{- 0.02}$ \\
        \\[-.5em]
        Ar/H & (Ar$^{2+}$ + Ar$^{3+}$)/H$^+$        & ---      &$ 1.01\pm0.01$ & $ 1.01\pm 0.01$ \\
        \\[-.5em]
        Ar/O & Ar$^{2+}$/(O$^+$ + O$^{2+}$)         & $1.36^{+ 0.66}_{-0.58}$      &$ 1.29\pm 0.02$ & $ 1.32^{+ 0.08}_{- 0.09}$ \\
        \\[-.5em]
        Kr/H & Kr$^{3+}$/H$^+$                            & $2.96^{+ 0.26}_{-0.23}$     & ---  & ---     \\
        \\
        \hline

    \end{tabular}
\begin{description}
\item $^{\rm a}$ {In their Tab. 3, \citet{delgadoingladaetal14} wrongly parameterise their ICF$_{\mathrm f}$ using $\upsilon$ while it should be using $\omega$, as they correctly write in their text.}
\end{description}
\end{table*}

\begin{table*}
    \centering
    \caption{Comparison between the elemental abundances determined in this work, using both the classical ICF schemes and ICFs obtained with machine-learning techniques, and those obtained in the literature.}
    \label{tab:total_ab}
    \begin{tabular}{lcccccc}
        \hline
        Element         &  ICF    & literature$^{\rm b}$         & look-up table & ANN       & \citet{liuetal00} & \citet{mcnabbetal16} \\
        \hline
        \\[-.5em]
        He$^{\rm c}$                 &                 & 
        10.95$-$11.00 (w) & ---       & ---   & 11.14  & 11.12 \\
        \\[-.5em]        
                                     &  &
        11.48$-$11.58 (c) & ---       & ---   &      &   \\
        \\[-.5em]        
        C (RLs)$^{\rm d}$   & C$^{2+}$/O$^{2+}$                      & 10.22$^{+0.07}_{-0.08}$          & ---       & ---   & 9.40   & 9.46  \\[-.5em]
        \\[-.5em]
        N                   & N$^+$/H$^+$                            & ---            & 8.31$\pm$ 0.06      & 8.37$^{+0.06}_{-0.07}$  & 8.36   & 8.20  \\
        \\[-.5em]
                            & N$^+$/O$^+$                            & 8.35$+^{0.11}_{-0.12}$  & 8.36$\pm {0.09}$   &  8.37$+^{0.09}_{-0.08}$    &        &       \\
        \\[-.5em]
        O (RLs)$^{\rm d}$   & (O$^+$ + O$^{2+}$)/H$^+$               & $10.63\pm0.04$    & ---       & ---   & 9.66   & 9.51  \\
        \\[-.5em]
        O                   & (O$^+$ + O$^{2+}$)/H$^+$               & 8.70$^{+0.04}_{-0.05}$          & 8.69$^{+0.04}_{-0.03}$      & $8.70\pm0.04$  & 8.70   & 8.51  \\
        \\[-.5em]
        S                   & (S$^+$ + S$^{2+}$)/H$^+$               & ---            & $7.12\pm0.04$     & $7.11\pm0.03$  & 7.21   & 7.00  \\
        \\[-.5em]
                            & (S$^+$ + S$^{2+}$)/O$^+$               & 7.18$^{+0.08}_{-0.07}$           & 7.16$^{+0.13}_{-0.11}$      & 7.12$^{+0.05}_{-0.04}$  &        &       \\
        \\[-.5em]
                            & (S$^+$ + S$^{2+}$)/(O$^+$ + O$^{2+}$)  & ---            & $7.12\pm0.04$      & 7.13$^{+0.04}_{-0.03}$  &        &       \\
        \\[-.5em]
        Cl                  & Cl$^{2+}$/H$^+$                        & ---            & $5.41\pm0.04$      & $5.42\pm0.03$  & 5.62   & 5.77  \\
        \\[-.5em]
                            & (Cl$^{2+}$ + Cl$^{3+}$)/H$^+$          & ---            & $5.41\pm0.03$      & $5.41\pm0.03$  &        &       \\
        \\[-.5em]
                            & Cl$^{2+}$/O$^+$                        & $5.49\pm0.08$           & 5.45$^{+0.14}_{-0.12}$  & 5.42$^{+0.05}_{-0.04}$  &        &       \\
        \\[-.5em]
        Ar                  &                                        & $6.65\pm0.05^{\rm e}$ & ---      & ---   & 6.43   & 6.20  \\
        \\[-.5em]
                            & (Ar$^{2+}$ + Ar$^{3+}$)/H$^+$          & ---            & $6.65\pm0.03$  & $6.65\pm0.03$  &        &       \\
        \\[-.5em]
                            & (Ar$^{2+}$/(O$^+$ + O$^{2+}$)          & 6.68$^{+0.18}_{-0.25}$           & 6.65$^{+0.03}_{-0.04}$   & $6.67\pm0.04$  &        &       \\
        \\[-.5em]
        Kr                  & Kr$^{3+}$/H$^+$                        & 4.03$^{+0.04}_{-0.03}$   & ---       & ---   & ---    & ---   \\
        \\[-.5em]
        \hline

    \end{tabular}

\begin{description}
\item $^{\rm a}$ {Abundances in units of 12+log(X/H)}
\item $^{\rm b}$ {ICFs for C, N, O, S, Cl and Ar from \citet{delgadoingladaetal14}; ICF for Kr from \citet{sterlingetal15}}
\item $^{\rm c}$ {He/H = He$^+$/H$^+$ + He$^{2+}$/H$^+$ for the warm (w) and cold (c) components (see Sect.~\ref{sec:helium_integ}.}
\item $^{\rm d}$ {Only using classical ICFs (see text).}
\item $^{\rm e}$ {Ar/H = Ar$^{2+}$/H$^+$ + Ar$^{3+}$/H$^+$ + Ar$^{3+}$/H$^+$.}
\end{description}
\end{table*}

 In Table~\ref{tab:total_ab}, we present the elemental abundances obtained from the three methods applied in this work. A comparison between the ICFs obtained using different techniques results in elemental abundances that are in relatively good agreement for all cases. It is worth mentioning that although the He$^{2+}$/He fraction in each plasma component of the nebula is quite uncertain, its low value has almost no effect on the only ICF we have used from \citet{delgadoingladaetal14}, which is parameterised using this ratio (i.e. ICF(O$^+$ + O$^{2+}$)), minimising possible systematic effects on our determinations. In Fig.~\ref{fig:hist_abun} we show the Monte Carlo distributions obtained for the elemental abundances using the ICFs shown in Table~\ref{tab:icfs}; only the ICF distributions obtained using more than one method are shown. The different approaches for computing the ICFs (literature, look-up table and ANN are shown in purple, green, and orange, respectively). The vertical lines on each colour show the median of the distribution.

Regarding the largest discrepancies we have found, they mostly pertain to ICFs relative to oxygen that use O$^+$ as a proxy (i.e., ICF((N$^+$)/O$^+$), ICF((S$^+$ + S$^{2+}$)/O$^+$), and ICF(Cl$^{2+}$/O$^+$)). This is because O$^+$ is residual in this PN, and it was computed using the red \forb{O}{ii} $\lambda$7330+ lines, which are severely affected by recombination contribution and potentially by telluric emission. Hence, despite our careful consideration of all these effects, they could introduce systematic uncertainties to the O$^+$/H$^+$ ratio. However, these systematic uncertainties are mitigated in the approach we took with the look-up table and ANN methods. In contrast, in the case of nitrogen, if we use the classical ICF from \citet{peimbertcostero69}, which assumes ICF(N$^+$/O$^+$) = 1.0, we obtain 12 + log(N/H) = 8.29, which is in slightly worse agreement with the abundances obtained using the other methods. Nonetheless, it is not entirely clear which ICF best represents this correction for PN, as a bias could be introduced in the set of photoionisation models selected by \citet{delgadoingladaetal14} \citep[see][]{delgadoingladaetal15, amayoetal20}. However, The largest discrepancy is in the ICF(Ar$^{2+}$/(O$^+$ +O$^{2+}$)) where it is clear that ICF of \citet{delgadoingladaetal14} show a huge dispersion compared to the look-up table and ANN methods. This is due to the fact the \citet{delgadoingladaetal14} method does not use Ar$^{3+}$/Ar$^{2+}$ as a constraint.

\subsection{Oxygen content in the cold region}

Following the method described in \citet[][see their Eqs. 5 and 6]{2022Garcia-Rojas_mnras510}, we compute the mass of oxygen in the form of O$^+$ and O$^{++}$ contained in the cold region, relative to that contained in the warm region. These ratios depend on the values of the electron temperature and densities in both regions, but not on $\omega$. The results are presented in Table~\ref{tab:mass_frac_warm_cold}, and confirm previous results from \citet{2020Gomez-Llanos_mnra497}, \citetalias{richeretal22} and \citetalias{2022Garcia-Rojas_mnras510}: the amount of oxygen in the rich and cold region is of the same order of magnitude as that embedded in the ``classical'' warm nebula.

\begin{table}
    \caption{Physical parameters and mass fraction between the warm and cold regions for O$^+$ and O$^{2+}$.}
    \centering
    \begin{tabular}{cc}
    \hline
    & \ngc \\
    \hline
$T_e^w$ [K] & 8,300 \\
$T_e^c$ [K] & 2,000 \\
$n_e^w$ [cm$^{-3}$]  & 3,400 \\
$n_e^c$ [cm$^{-3}$]  & 10,000 \\
M$^c$/M$^w$(O$^+$) & 1.2 \\
M$^c$/M$^w$(O$^{2+}$) & 0.7 \\
\hline
    \end{tabular}
    
    \label{tab:mass_frac_warm_cold}
\end{table}

\section{Discussion}\label{sec:discuss}

\subsection{Comparison with previous works}

The most detailed chemical abundance studies in the literature for NGC\,6153 are those by \citet{liuetal00}, \citet{tsamisetal08}, \citet{mcnabbetal16} and \citetalias{richeretal22}. However, not all of these authors provide elemental abundances for this PN. \citet{tsamisetal08} used IFU spectra of a portion of NGC\,6153, covering from the central star to the south-eastern outskirts of the PN, and focused on computing the ADF(O$^{2+}$) and on finding correlations of the ADF with different physical parameters in the PN. They did manage to compute the elemental He abundance, yielding a value of 12 + log(He/H) = 11.12. On the other hand, \citetalias{richeretal22} conducted a detailed analysis of position-velocity diagrams of multiple emission lines from high-spectral resolution spectra of NGC\,6153, aiming to disentangle the kinematics of the gas. Although they also computed physical conditions, ionic abundances, and ADFs as a function of gas kinematics, they did not attempt to compute elemental abundances from their integrated spectra. Finally, despite being carried out with a high-quality data set, comparisons with the results by \citet{mcnabbetal16} should be made with extreme caution because of their extremely low \ha/\hb ratio ($\sim$1.0), which clearly reveals problems with data reduction. 

In Table~\ref{tab:total_ab} we show, for comparison, the abundances reported in \citet{liuetal00} and \citet{mcnabbetal16}. It is remarkable the excellent agreement between O abundances from CELs obtained in this work and by \citet{liuetal00}. This good agreement extends, albeit to a lesser extent, to abundances of other species such as N and S (less than $\sim$0.1 dex differences). Conversely, the differences between the abundances obtained with recombination lines are enormous, covering almost an order of magnitude in C (0.87 dex) and O (0.98 dex). 

On the other hand, the agreement with the elemental abundances obtained by \citet{mcnabbetal16} is relatively poor, with differences ranging from 0.06 dex to 0.23 dex in abundances obtained from CELs. Particularly puzzling is the difference in the O abundance, which amounts to 0.19 dex. This difference cannot be attributed to a different temperature scheme, as we have assumed as representative for O$^{2+}$, the dominant ion of this species, a \Te which is only $\sim$250\,K lower than that assumed by these authors. However, we have found that the fluxes of the [O~{\sc iii}] $\lambda\lambda$4959,5007 lines relative to \hb reported by these authors are between 49--51 per cent lower than those reported in this work or in \cite{liuetal00}. One could invoke differences in the excitation of the nebula in the different volumes studied as the cause of this behaviour. However, similar differences (and in the same direction) are found in the fluxes of the \forbl{O}{ii}{7320+30} lines, suggesting a systematic problem with the fluxes reported by these authors that invalidates any attempt at comparison with their results. 

\subsection{The effect of having two different plasma components in a photoionised nebula}\label{sec:}

As outlined in Sect.~\ref{sec:intro}, the hypothesis of several plasma components with significantly different physical conditions in PNe was first proposed by \citet{torrespeimbertetal90}, and since then, it has been commonly invoked as an explanation for the discrepancy found between abundances obtained from faint optical CNONe ionic RLs and the corresponding bright CELs of the same ions. However, it has only been very recently discovered that metal RLs emit from a plasma component that is clearly distinct from the CEL emission region in a small group of high-ADF PNe.

In an outstanding study using high-spectral resolution data of NGC\,6153, \citetalias{richeretal22} were able to disentangle a complex temperature structure in the plasma and two different kinematic components: the kinematics of most emission lines, including H and He RLs and all CELs followed a classical expansion law with outer parts of the nebula expanding faster, while heavy element RLs followed a constant expansion velocity, defining an additional plasma component. However, those authors also found, from the comparison of \Te({\forb{O}{iii}}) and \Te(\forb{Ar}{iii}), that a contribution of relatively large temperature fluctuations in the warm component of the gas to the measured ADF cannot be ruled out (see their Sect. 4.1 and Fig. 40). This behaviour complicates the computation of the real abundances even more, especially in the warm component, as in such a case, the derived chemical abundances from CELs would be severely underestimated \citep[see e.g.][]{2023Mendez-Delgado_Natur618}. 

\citet{2009Bohigas_rmxaa45} and \citet{2015Bohigas_mnras453} proposed studying the effect on abundance determinations of weighting the two different plasma emitting regions analytically by comparing observational data with relatively simple two-phase photoionisation models. 
In this paper, we explore a situation where different types of metal lines (collisional and recombination) are mainly emitted by two different regions or gas phases. This implies that the \hb intensity, which is usually used to normalise the metal lines, needs to be distributed between the two components (this is the role of $\omega$). The classical derivation of the ADF, defined as the ratio of line intensities and emissivities as outlined by \citet{2009Bohigas_rmxaa45} in its equations 1 and 2, is based on the simplification that \hb emission is common to both sources. This is the difference between the ADF and the ACF.

Recently, a detailed photoionisation modelling of NGC\,6153 was carried out by \citet{2020Gomez-Llanos_mnra497}. These authors found that a given ADF(O$^{2+}$) could be reproduced by different combinations of ``normal'' and metal-rich components, each with different abundance contrast factors. This introduces a degeneracy in the ADF-ACF relationship, potentially causing apparently low-ADF PNe to conceal two distinct plasma components and a high ACF. Resolving this degeneracy is crucial to untangle the abundance discrepancy puzzle. Additionally, these authors stated that very high spatial and/or spectral resolution observations were the only means to isolate the contribution of the region from the total H\textsc{i} emission and resolve this important issue. We have made a preliminary attempt to isolate this contribution by combining our data with that of \citetalias{richeretal22}. However, as these previous authors wisely pointed out in their work, reality is complex. Nonetheless, we believe that our work represents a foundational step towards a deeper understanding of this type of PN. The key lies in acquiring data that facilitate reliable measurements of electron temperatures in both the cold and warm components, as well as the temperature of H\textsc{i} through the Balmer and/or Paschen decrements. These are all necessary observables for computing $\omega$ and, consequently, the ACF.

The study of spatially resolved data of PNe to address the abundance discrepancy problem has been proposed in several key projects to be developed for several upcoming IFU facilities at medium- to large-sized telescopes (4-10m). For instance, the Mirror-slicer Array for Astronomical Transients \citep[MAAT; see Sect. 3.6 in][]{2020Prada_arXiv} is intended to be attached to the Gran Telescopio Canarias (GTC), the WEAVE Stellar, Circumstellar and Interstellar Physics survey \citep[SCIP; see Sect. 4.2 in]{2023Jin_mnras715} at the 4.2m {\it William Herschel} Telescope, and the future Wide-field Spectroscopic Telescope \citep[WST; see Sect. 3.5.3 in]{2024Mainieri_arXiv} would significantly increase the amount of this type of data in the coming years. Therefore, it is necessary to establish the basis for a correct analysis and interpretation of the data, which is the aim of this work.

Based on the values presented in Tab.~\ref{tab:mass_frac_warm_cold}, we can calculate the product \Ne$\cdot$\Te, which is associated with the gas pressure. The pressure in the warm component seems to be comparable to that in the cold region (only 40\% higher). Due to the uncertainties in determining the density of the cold region, we cannot definitively say if the two phases are in pressure equilibrium. This apparent lack of pressure equilibrium is common in this type of object, as the difference in density between both components is not as large as the difference in temperatures. There is at present no explanation of this fact. It is possible that the two plasma components might not be in contact because they originate from two different ejections of gas at different evolutionary times.

Finally, it is worth mentioning that very recently, observational results from high-quality observations of {\hii} regions have revealed that the temperature fluctuations scenario applies to {\hii} regions in the high-ionised volume of the gas \citep{2023Mendez-Delgado_Natur618}. However, no evidence of such behaviour was found for PNe, although their conclusions might be affected by the small sample size, the limited metallicity range covered by the high-quality data of PNe, and the inclusion of several high-ADF PNe in the sample, which were analysed in the same way as low-ADF PNe. A more detailed analysis of an extended sample of high-quality PN spectra from the literature, with high S/N ratio detections of the weak \perm{O}{ii} RLs, as well as a careful analysis of the recombination contamination to the auroral lines, could unravel the role of possible temperature fluctuations in the observed abundance discrepancy. This sample should be free of known high-ADF PNe as well as known post-common-envelope systems harbouring a close binary central star, which could alter the conclusions of the analysis.

\section{Summary}\label{sec:summary}

In this work, we present deep integral-field unit (IFU) spectroscopy of the high ADF (ADF $\sim$10) PN NGC\,6153. The spectra were obtained with the MUSE spectrograph covering the wavelength range 4600$-$9300 \AA\ with effective spectral resolution from R = 1,609 to R = 3,506 for the bluest to the reddest wavelengths, respectively, and with a spatial sampling of 0.2 arcsec.

We built spatially resolved maps for 60 emission lines, as well as for two continuum regions bluer and redder to the \hi\ Paschen discontinuity. We produced a spatially resolved map of a neutron-capture emission line, \forbl{Kr}{iv}{5867.74}, which, owing to the faintness of this line, has only been attempted in one other PN.

We followed the methodology described in \citetalias{2022Garcia-Rojas_mnras510} to perform the emission line analysis. Extinction correction is determined pixel by pixel, showing spatial variations indicative of local extinction. Accounting for a cold plasma phase could affect extinction corrections, but owing to the relatively small effect of this correction in the computed c(\hb) and to avoid adding noise to our data, we skipped applying any correction in extinction computations for the presence of the cold region. Correction for recombination contributions to \forb{N}{ii} and \forb{O}{ii} auroral lines is performed following the methodology described in \citetalias{2022Garcia-Rojas_mnras510}. These corrections led to very different \Te(\forb{N}{ii}) maps, highlighting the importance of accounting for multiple plasma phases in accurate nebular analyses.

The electron temperature (\Te) and density (\Ne) from CELs were determined using the methodology described in \citetalias{2022Garcia-Rojas_mnras510} and atomic data from Table~\ref{tab:atomic_data}. Electron temperatures were also obtained from RL diagnostics, as \hei RLs and the {\hi} Paschen jump (PJ), which are significantly lower than those obtained using CEL diagnostics, pointing to the presence of a low temperature plasma component. 

Ionic chemical abundance maps were constructed using emission lines with high signal-to-noise ratios. We used a two-ionisation zone scheme to compute line emissivities. The He$^+$/H$^+$ and He$^{2+}$/H$^+$ ratios are determined separately for the warm and cold components, using the method described by \citet{2023Morisset_arXiv}.

By using the \Te(\perm{O}{ii}) determined for NGC\,6153 by \citetalias{richeretal22} we could estimate the contribution of the cold region in the \hb emission, and apply it to chemical abundance calculations. Maps of $\omega$, the weight of the cold component, were generated, and correction factors were applied to the ionic abundances computed from both CELs and RLs. Maps of the abundance contrast factor (ACF) were also produced to account for the presence of the cold region. Ionic abundance maps taking into account this correction were presented. 

We computed ionic abundances from the integrated spectrum of NGC,6153 following different assumptions on the \Te structure and considering or ignoring the weight of the cold component in the \hb emission. The ionic abundances obtained in this way from CELs are only slightly modified (an average of $\sim$0.05 dex higher) with respect to a classical analysis; however, ionic abundances from RLs are nearly an order of magnitude higher.

We selected as representative for the integrated spectrum those ionic abundances computed assuming a three-zone ionisation scheme, where each ion emissivity (including H$^+$) was computed using its corresponding physical conditions, and considering the $\omega$ factor. Ionisation correction factors (ICFs) were calculated using various methods: ICFs from the literature, look-up tables from photoionisation models, and ad hoc ICFs using machine learning techniques. In general, the three methods provide very consistent ICFs among each other. However, higher differences were found in ICFs relative to O, which use O$^+$ as a proxy, owing to the relatively residual O$^+$/H$^+$ abundance in this PN and to potential systematic uncertainties in O$^+$/H$^+$ ratio determinations, which were derived from the red \forbl{O}{ii}{7320+30} lines which are severely affected by recombination contribution and potentially affected by telluric emission. Elemental abundances in high-excitation PNe should therefore be taken with caution, especially for elements requiring a large ICF, as is the case of nitrogen.

While an observational connection between the observed abundance discrepancy and temperature fluctuations in {\hii} regions has recently emerged, the situation regarding PNe remains enigmatic. In recent years, several studies have illustrated the presence of at least two distinct plasma components in certain PNe, often linked to short-period binary central stars that have undergone a common envelope phase. However, a comprehensive explanation for the abundance discrepancy in PNe remains elusive. We cannot definitively discard the possibility that temperature fluctuations in the warm component of the gas may partially contribute to the observed abundance discrepancy. If so, the chemical analysis of high-ADF objects would become much more complex, as ionic abundances derived from CELs may no longer be valid. Only a thorough and coherent analysis of a large sample of PN spectra covering a wide range of parameters (metallicities, excitation conditions, etc.) will shed light on this matter.

\section*{Acknowledgements}
We want to dedicate this paper to the memory of our dear colleague and friend Claudio Mendoza Guardia (1951-2024) who passed away during the preparation of this paper. His contributions to atomic data computations for nebular purposes have been invaluable to the field. His insistence also made it possible for proper credit to be given to the producers of atomic data in nebular works, as it is usually done in the last years. We sincerely thank Gra\.{z}yna Stasi\'nska, the referee of this paper, for her valuable comments that enhanced its quality.
This paper is based on observations made with ESO Telescopes at the Paranal Observatory under program ID 097.D-0241. VG-LL, JG-R and DJ acknowledge financial support from the Canarian Agency for Research, Innovation, and Information Society (ACIISI), of the Canary Islands Government, and the European Regional Development Fund (ERDF), under grant with reference ProID2021010074, from the Agencia Estatal de Investigaci\'on of the Ministerio de Ciencia e Innovaci\'on (AEI- MCINN) under grants ``Espectroscop\'ia de campo integral de regiones H II locales. Modelos para el estudio de regiones H II extragala\'acticas'' with reference DOI:10.13039/501100011033, and ``Nebulosas planetarias como clave para comprender la evoluci\'on de estrellas binarias'' with reference PID-2022-136653NA-I00 (DOI:10.13039/501100011033). We also acknowledge support from project P/308614 financed by funds transferred from the Spanish Ministry of Science, Innovation and Universities, charged to the General State Budgets and with funds transferred from the General Budgets of the Autonomous Community of the Canary Islands by the MCINN. JG-R also acknowledges funds from the AEI-MCINN, under Severo Ochoa Centres of Excellence Programme 2020-2023 (CEX2019-000920-S). DJ also acknowledges support from the AEI-MCINU under grant ``Revolucionando el conocimiento de la evoluci\'on de estrellas poco masivas'' and the European Union NextGenerationEU/PRTR with reference CNS2023-143910 (DOI:10.13039/501100011033). 
CM acknowledges funds from UNAM/DGAPA/PAPIIT IN 101220 and IG 101223. RW acknowledges support from STFC Consolidated grant ST/W000830/1. This work has made use of the computing facilities available at the Laboratory of Computational Astrophysics of the Universidade Federal de Itajub\'a (LAC-UNIFEI). The LAC-UNIFEI is maintained with grants from CAPES, CNPq and FAPEMIG.

\section*{DATA AVAILABILITY}

The data products used in this paper: raw and reduced MUSE data cubes and extracted emission line maps, are available from the ESO archive facility at \url{http://archive.eso.org/}. The analysis pipeline, the emission line maps and the {\sc python} scripts used for the analysis and to produce the tables and figures presented in this paper are available at \url{https://github.com/VGomezLlanos/NGC-MUSE}.






 \newpage
\begin{appendix}

\section{He$^+$/H$^+$ abundance maps}
\label{sec:ionic_ab_maps_he}

In this section, we discuss the case of He lines, which can be emitted by both cold and warm components.
This is the general case described by \citet{2023Morisset_arXiv}:

\begin{equation}
\begin{split}
\frac{I_\lambda}{I_\beta} = & \left(\frac{X^i}{H^+}\right)^w \cdot (1-\omega) \cdot \frac{\epsilon_\lambda(T_{\rm e}^w, n_{\rm e}^w)}{\epsilon_\beta(T_{\rm e}^w, n_{\rm e}^w)} + \\
                            & \left(\frac{X^i}{H^+}\right)^c  \cdot \omega \cdot \frac{\epsilon_\lambda(T_{\rm e}^c, n_{\rm e}^c)}{\epsilon_\beta(T_{\rm e}^c, n_{\rm e}^c)}.
\end{split}
\label{eq:ion_ab_2comp}
\end{equation}

Using two \hei emission lines with different temperature dependencies (namely \alloa{He}{i}{6678} and \alloa{He}{i}{7281}), we can determine the contribution of each component to the total emission and derive the ionic abundance in each component:

\begin{equation}
    \left(\frac{He^+}{H^+}\right)^w = \frac{\epsilon_\beta^w}{(1-\omega) \cdot I_\beta} \cdot \frac{I_{6678} - I_{7281} \cdot \frac{\epsilon_{6678}^c}{\epsilon_{7281}^c}}{\epsilon_{6678}^w - \epsilon_{7281}^w \cdot \frac{\epsilon_{6678}^c}{\epsilon_{7281}^c}},
\label{eq:he1warm}
\end{equation}
and
\begin{equation}
    \left(\frac{He^+}{H^+}\right)^c = \frac{\epsilon_\beta^c}{\omega \cdot I_\beta} \cdot \frac{I_{6678} - I_{7281} \cdot \frac{\epsilon_{6678}^w}{\epsilon_{7281}^w}}{\epsilon_{6678}^c - \epsilon_{7281}^c \cdot \frac{\epsilon_{6678}^w}{\epsilon_{7281}^w}},
\label{eq:he1cold}
\end{equation}

\begin{figure*}
    \centering
    \includegraphics[width=17cm]{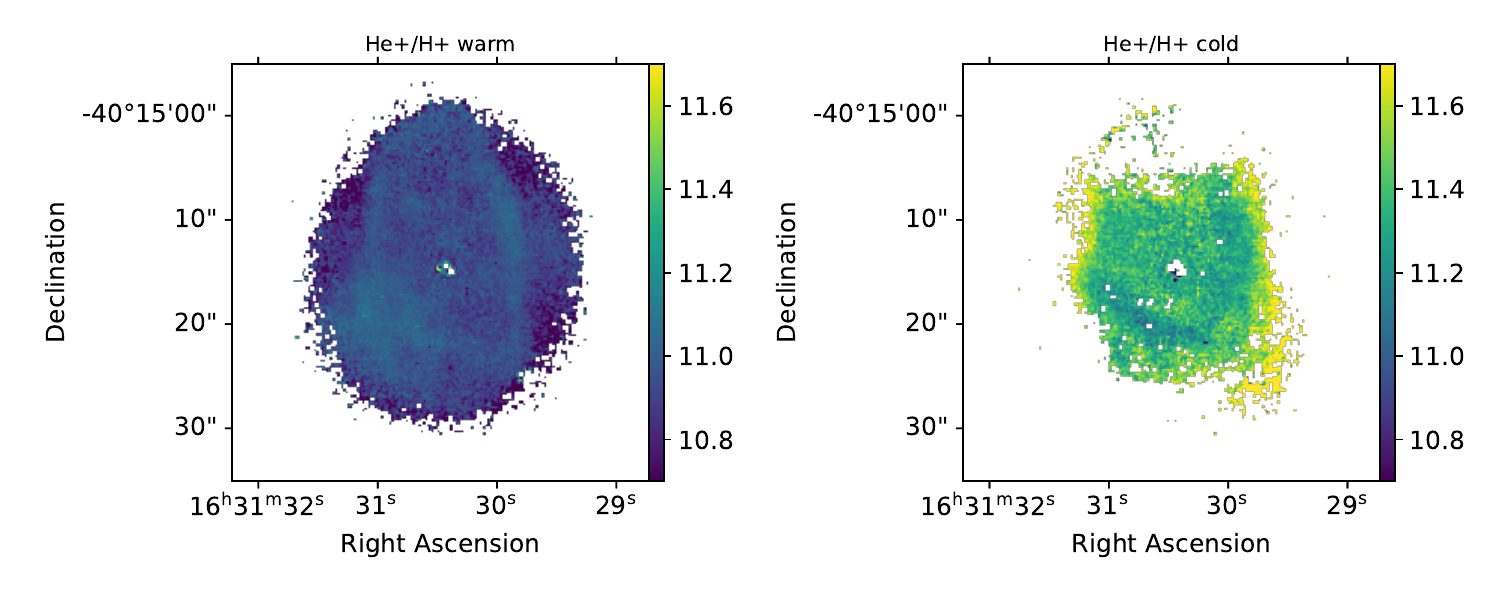}
    \caption{He$^+$/H$^+$ ionic abundance maps for the warm and cold region in the left and right panel, respectively. }
    \label{fig:he_ab_map}
\end{figure*}

The physical parameters for the warm (cold) component are as follows: \Te= 8,300 (2,000)\,K, and \Ne=3,400 (10,000)\,cm$^{-3}$. 
The ionic abundance maps for He$^+$/H$^+$ in each component are shown in Fig.~\ref{fig:he_ab_map}. The He$^+$ ion is clearly more abundant in the metal-rich region, but with an enhancement of the order of 0.5 dex, smaller than the 2 dex overabundance observed for the metals.

For the He$^{2+}$/H$^+$ ratio, the situation is less simple since only one line of the residual \heii ion is observed. We do not try to compute any value for the abundance of He$^{2+}$/H$^+$ (but see Sec~\ref{sec:helium_integ} for the case of the integrated spectrum). 

\section{\Te-\Ne recipes to compute ionic abundances}
\label{sec:recipes}

In this section we describe the different \Te-\Ne recipes explored in this work to compute ionic abundances.

\begin{itemize}
    \item Recipe 1: The most simple case is to adopt a general ionisation scheme with 2 zones, where {\forbr{N}{ii}{5755}{6548}$-$\forbr{S}{ii}{6716}{6731}} are the physical conditions for ions with IP<17~eV, and {\forbr{S}{iii}{6312}{9069}$-$\forbr{Cl}{iii}{5518}{5538}} for ions with IP$\geq$17~eV. These conditions are used to compute the emissivities from CELs, RLs, and the \hb line. 

    \item Recipe 2: The abundances are obtained from CELs using the same 2-IP zones as in Recipe 1. For heavy element abundances from RLs, we adopt \Te= 2,000~K and \Ne= 10,000~cm$^{-3}$ for both the line and \hb. This is the recipe we used to generate ADF maps (left panels of Fig.~\ref{fig:adf_op_opp}).

    \item Recipe 3: We follow a similar scheme as in Recipe 2, but we use the Paschen jump temperature to derive the \hb emissivity in the determination of the ionic abundances, by setting the \verb|tem_HI| keyword when calling the \verb|Atom.getIonAbundance| method of \pyneb. 
    
This last recipe would be the most complex one in an effort to compute the emissivities of each ion (including H$^+$) using its corresponding temperature. But in the hypothesis of two separate plasma of very different chemical compositions, the H$^+$ emission needs to be split into its two corresponding contributions, as in the next recipes where we take $\omega$ into account following Eqs.~\ref{eq:ion_ab_w} and \ref{eq:ion_ab_c}. 

    \item Recipes 4, 5 and 6: The abundances from the RLs are obtained as in Recipe 2, but multiplied by $1/\omega$ as in Eq.~\ref{eq:ion_ab_c}. For the CELs, the abundances are multiplied by $1/(1-\omega)$ as in Eq.~\ref{eq:ion_ab_w}. An additional high IP zone is defined for ions with IP>35 eV. The corresponding abundances are obtained considering \Te(\forb{S}{iii})-\Ne(\forb{Cl}{iii}), \Te(\forb{Ar}{iv})-\Ne(\forb{Ar}{iv}) and the weighted average of \Te(\forb{Ar}{iii}) and \Te(\forb{Ar}{iv}) given by Eq.~\ref{eq:ar_average} with \Ne(\forb{Ar}{iv}) for Recipes 4, 5 and 6 respectively. Indeed, recipe 4 was the one used to generate the ionic abundance maps shown in Fig.~\ref{fig:ionic_ab_omega} and the ACF maps shown in the right panels of Fig.~\ref{fig:adf_op_opp}.
    
    \item Recipe 7: The same scheme as in Recipe 6 is used, the only difference being the use of \Te = 8,300~K for the \hb emissivity when computing the abundances of CELs. 

\end{itemize}

\section{Emission line flux maps and line fluxes table for the integrated spectra}
\label{sec:app:lines}

In this section, we show the derredened flux maps for emission lines of 8 elements heavier than H at different ionisation stages, as well as the table with the observed and derredened fluxes of 60 emission lines in the integrated spectra of NGC\,6153.

\begin{figure*}
    \centering
    \includegraphics[scale = 0.69]{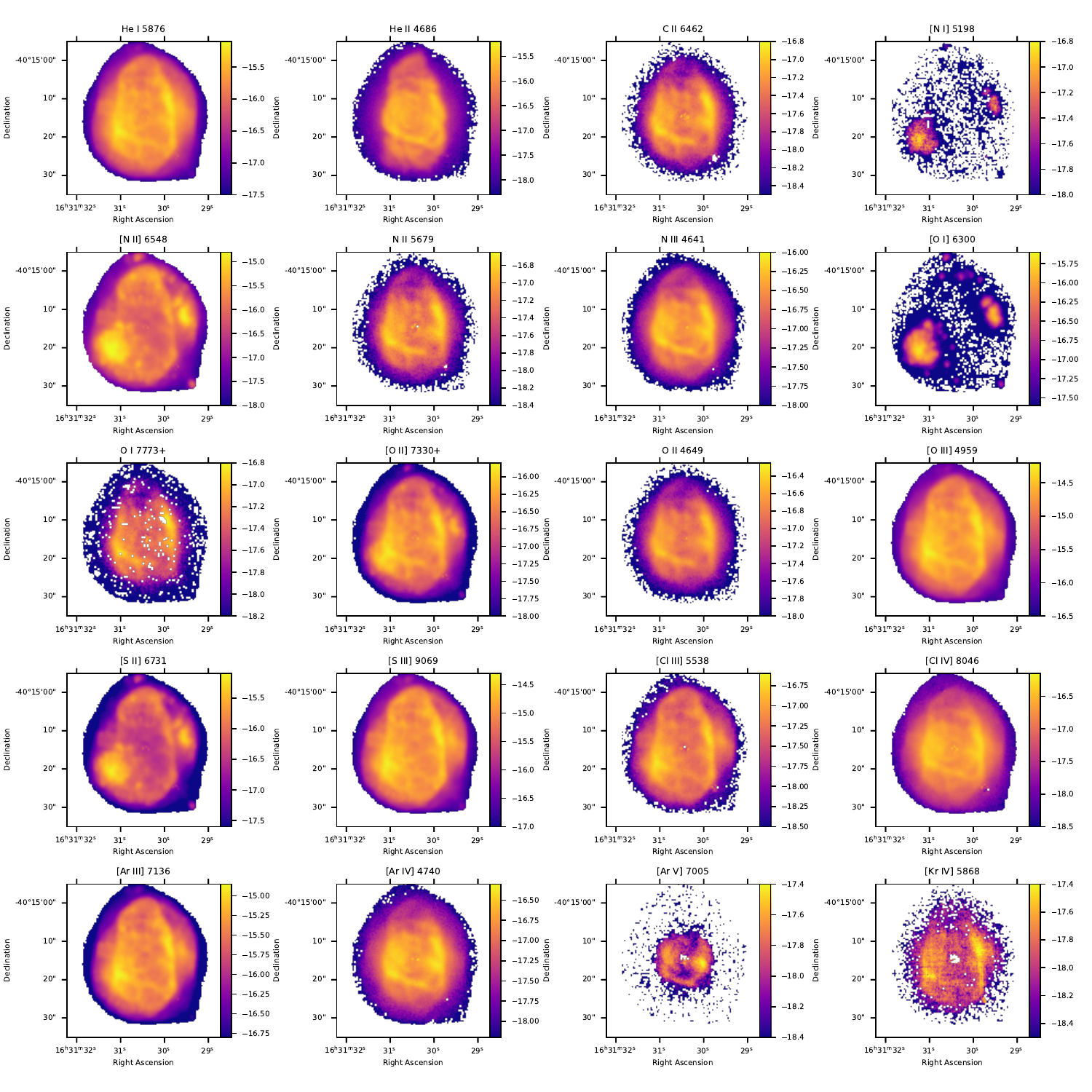}
    \caption{Flux maps of the observed emission lines of eight elements (heavier than H) in different ionisation states. The data were obtained from the MUSE datacube of \ngc. The flux is in units of erg cm$^{-2}$s$^{-1}$\AA$^{-1}$ and the maps are classified by the atomic mass and the ionisation potential of the ion. We do not show here the special case of the auroral line \forbl{N}{ii}{5755} which is contaminated by recombination and is discussed in more detail in Sect~\ref{sec:corr_rec}}.
    \label{fig:line_fluxes}
\end{figure*}

\begin{table}[htb]
    \centering
    \scriptsize
    \caption{Measured (F) and derredened (I) line fluxes in the integrated spectra. }
    \label{tab:int_fluxes}
    \begin{tabular}{llccl}
        \hline
        Line   & $\lambda_0$(\AA) & F($\lambda$) & I($\lambda$) & Notes \\
        \hline
        \perm{O}{ii} & 4641.81 &      3.6 $\pm$    0.2 &      4.3 $\pm$    0.3 & blend with \perm{N}{iii} $\lambda$4641.85 \\
\perm{O}{ii} & 4649.13    &     1.85 $\pm$   0.09 &     2.18 $\pm$   0.16 & blend with \perm{O}{ii} $\lambda$4650.85  \\
\forb{Fe}{iii} & 4658.17  &    0.068 $\pm$  0.004 &    0.080 $\pm$  0.006 &   \\
\perm{O}{ii} & 4661.63  &     0.42 $\pm$   0.02 &     0.49 $\pm$   0.04 &   \\
\perm{He}{ii} & 4685.71 &     11.6 $\pm$    0.6 &     13.4 $\pm$    0.9 &   \\
\forb{Ar}{iv} & 4711.37  &      1.5 $\pm$    0.1 &      1.7 $\pm$    0.2 &   \\
\perm{He}{i} & 4713.14 &     1.27 $\pm$   0.09 &     1.42 $\pm$   0.13 &   \\
\forb{Ne}{iv} & 4724.15   &    0.022 $\pm$  0.001 &    0.025 $\pm$  0.002 & blend with \forb{Ne}{iv} $\lambda$4625.85  \\
\forb{Ar}{iv} & 4740.17  &      2.2 $\pm$    0.1 &      2.4 $\pm$    0.2 &   \\
\perm{N}{ii} & 4802.79 &     0.13 $\pm$   0.01 &     0.14 $\pm$   0.01 &   \\
\perm{H}{i} & 4861.32  &    100.0 $\pm$    5.0 &    100.0 $\pm$    5.0 &    \\
\perm{He}{i} & 4921.93  &     1.82 $\pm$   0.09 &     1.73 $\pm$   0.12 &    \\
\forb{O}{iii} & 4958.91  &    316.5 $\pm$   15.8 &    293.2 $\pm$   20.1 &    \\
\forb{Ar}{iii} & 5191.82  &    0.088 $\pm$  0.004 &    0.069 $\pm$  0.005 &    \\
\forb{N}{i} & 5197.90  &     0.23 $\pm$   0.01 &     0.17 $\pm$   0.01 &    \\
\forb{N}{i} & 5200.26  &    0.055 $\pm$  0.003 &    0.042 $\pm$  0.003 &    \\
\perm{C}{ii} & 5342.38  &    0.155 $\pm$  0.008 &    0.108 $\pm$  0.007 &    \\
\forb{Kr}{iv} & 5346.02 &    0.029 $\pm$  0.002 &    0.020 $\pm$  0.001 &    \\
\perm{He}{ii} & 5411.52 &     1.51 $\pm$   0.08 &     1.01 $\pm$   0.06 & blend with \forb{Fe}{iii} $\lambda$5412.00   \\
\forb{Cl}{iii} & 5517.71  &     0.89 $\pm$   0.04 &     0.56 $\pm$   0.03 &    \\
\forb{Cl}{iii} & 5537.88  &     1.07 $\pm$   0.05 &     0.66 $\pm$   0.04 &    \\
\perm{N}{ii} & 5666.63  &     0.36 $\pm$   0.02 &     0.21 $\pm$   0.01 &    \\
\perm{N}{ii} & 5676.02  &    0.164 $\pm$  0.008 &    0.094 $\pm$  0.006 &    \\
\perm{N}{ii} & 5679.56  &     0.83 $\pm$   0.04 &     0.47 $\pm$   0.03 &    \\
\perm{N}{ii} & 5686.21  &    0.123 $\pm$  0.006 &    0.070 $\pm$  0.004 &    \\
\perm{N}{ii} & 5710.77  &    0.126 $\pm$  0.006 &    0.071 $\pm$  0.004 &    \\
\forb{N}{ii} & 5754.64  &     1.55 $\pm$   0.08 &     0.51 $\pm$   0.05$^{\rm a}$ &    \\
\forb{Kr}{iv} & 5867.74 &    0.097 $\pm$  0.005 &    0.050 $\pm$  0.003 &    \\
\perm{He}{i} & 5875.66  &     36.6 $\pm$    1.8 &     18.8 $\pm$    1.1 &    \\
\forb{O}{i} & 6300.30  &     1.96 $\pm$   0.10 &     0.82 $\pm$   0.05 &    \\
\forb{S}{iii} & 6312.10  &      3.3 $\pm$    0.2 &      1.4 $\pm$    0.1 &    \\
\forb{O}{i} & 6363.78  &     0.64 $\pm$   0.03 &     0.26 $\pm$   0.02 &    \\
\perm{C}{ii} & 6461.95  &     0.59 $\pm$   0.03 &     0.23 $\pm$   0.01 &    \\
\forb{N}{ii} & 6548.04  &     39.9 $\pm$    2.0 &     15.0 $\pm$    0.9 &    \\
\perm{H}{i} & 6562.80  &    735.1 $\pm$   36.8 &    274.4 $\pm$   14.6 &    \\
\perm{C}{ii} & 6578.05  &     1.64 $\pm$   0.08 &     0.61 $\pm$   0.04 &    \\
\forb{N}{ii} & 6583.45  &    125.1 $\pm$    6.3 &     46.3 $\pm$    2.7 &    \\
\perm{He}{i} & 6678.15  &     14.2 $\pm$    0.7 &      5.1 $\pm$    0.3 &    \\
\forb{S}{ii} & 6716.44  &      9.6 $\pm$    0.5 &      3.4 $\pm$    0.2 &    \\
\forb{S}{ii} & 6730.82  &     15.9 $\pm$    0.8 &      5.5 $\pm$    0.3 &    \\
\forb{Ar}{v} & 7005.40 &    0.068 $\pm$  0.003 &    0.021 $\pm$  0.001 &    \\
\perm{He}{i} & 7065.22 &     15.7 $\pm$    0.8 &      4.8 $\pm$    0.3 &    \\
\forb{Ar}{iii} & 7135.80  &     71.1 $\pm$    3.6 &     21.2 $\pm$    1.3 &    \\
\forb{Ar}{iv} & 7170.50  &    0.223 $\pm$  0.011 &    0.066 $\pm$  0.004 &    \\
\perm{He}{ii} & 7177.52  &    0.485 $\pm$  0.024 &    0.142 $\pm$  0.008 &    \\
\perm{C}{ii} & 7231.33  &      3.3 $\pm$    0.2 &      1.0 $\pm$    0.1 & blend with \perm{C}{ii} $\lambda\lambda$7236.42,7237.17   \\
\forb{Ar}{iv} & 7262.70  &    0.152 $\pm$  0.008 &    0.043 $\pm$  0.003 &    \\
\perm{He}{i} & 7281.35  &      2.5 $\pm$    0.1 &      0.7 $\pm$    0.0 &    \\
\forb{O}{ii} & 7318.92 &      7.8 $\pm$    0.4 &      1.1 $\pm$    0.1$^{\rm a}$ & blend with \forb{O}{ii} $\lambda$7319.99   \\
\forb{O}{ii} & 7329.67 &      6.3 $\pm$    0.3 &      0.9 $\pm$    0.1$^{\rm a}$ & blend with \forb{O}{ii} $\lambda$7330.73   \\
\forb{Cl}{iv} & 7530.54  &     1.00 $\pm$   0.05 &     0.26 $\pm$   0.02 &    \\
\forb{Ar}{iii} & 7751.10  &     20.7 $\pm$    1.0 &      5.0 $\pm$    0.3 &    \\
\perm{O}{i} & 7771.94   &     0.57 $\pm$   0.04 &     0.14 $\pm$   0.01 & blend with \perm{O}{i} $\lambda\lambda$ 7774.17,7775.39   \\
\forb{Cl}{iv} & 8045.62  &      2.6 $\pm$    0.1 &      0.6 $\pm$    0.0 &    \\
\perm{H}{i} & 8100    &    0.276 $\pm$  0.014 &    0.061 $\pm$  0.004 &  Blue continuum PJ  \\
\perm{C}{iii} & 8196.48   &    0.589 $\pm$  0.029 &    0.126 $\pm$  0.008 &  \\
\perm{H}{i} & 8400    &    0.114 $\pm$  0.006 &    0.023 $\pm$  0.001 &  Red continuum PJ  \\
\perm{H}{i} & 8750.47  &      6.1 $\pm$    0.3 &      1.1 $\pm$    0.1 &    \\
\perm{H}{i} & 8862.78  &      7.8 $\pm$    0.4 &      1.4 $\pm$    0.1 &    \\
\perm{H}{i} & 9014.91  &      9.0 $\pm$    0.5 &      1.6 $\pm$    0.1 &    \\
\forb{S}{iii} & 9068.60  &    202.7 $\pm$   10.1 &     35.3 $\pm$    2.3 &    \\
\perm{H}{i} & 9229.01  &     15.4 $\pm$    0.8 &      2.6 $\pm$    0.1 &    \\
\hline
    \end{tabular}
\begin{description}
\item $^{\rm a}$ {Derredened fluxes corrected from recombination contribution}
\end{description}
\end{table}

\end{appendix}

\end{document}